\documentclass[preprint,12pt]{elsarticle}

\usepackage[figuresright]{rotating}

\usepackage{graphicx}
\usepackage{dcolumn}
\usepackage{bm}
\usepackage{multirow}

\usepackage[colorlinks,linkcolor=blue,citecolor=blue]{hyperref}
\usepackage{amsmath,amssymb}
\usepackage{subfigure}
\usepackage{caption}    
\usepackage{color}
\usepackage{url}
\usepackage{lineno}
\modulolinenumbers[5]
\journal{Journal of Computational Physics}

\begin{document}

\begin{frontmatter}



\title{A two stage fourth-order gas-kinetic scheme on unstructured hybrid mesh}

\author{Dongxin Pan}
\ead{chungou@mail.nwpu.edu.cn}

\author{Chengwen Zhong\corref{cor1}}%
\ead{zhongcw@nwpu.edu.cn}

\author{Congshan Zhuo\corref{}}%
\ead{zhuocs@nwpu.edu.cn}
\cortext[cor1]{Corresponding author}
\address{National Key Laboratory of Science and Technology on Aerodynamic Design and Research, Northwestern Polytechnical University, Xi'an, Shaanxi 710072, China.}

\date{\today}

\begin{abstract}
This paper presents an accurate and robust fourth order gas-kinetic scheme on two dimensional unstructured hybrid mesh for incompressible and compressible viscous flows. For generalized Riemann problem and Navier-Stokes solution, the gas-kinetic scheme (GKS) provides a time-accurate flux solver using a different way in the reconstruction at a cell interface in which two slopes for the equilibrium state are used. Different from the previous one-stage time-stepping method, the two-stage Lax-Wendroff type time stepping method is applied in this paper. Compared to standard four-stage fourth-order Runge-Kutta method, the two-stage fourth order time accurate method reduces the complexity of the adoption of time derivative of the flux function. To achieve fourth order accuracy, a finite volume method for GKS using cubic spline reconstruction is proposed on both structured grid and unstructured hybrid mesh. When dealing with flow discontinuities, the original spline scheme is replaced by the one blended with shock-capturing WENO scheme. Many one and two-dimensional test cases, including Couette flow, Shu-Osher problem, Woodward-Colella blast problem,  two-dimensional Riemann problem,  viscous shock tube flow, supersonic flow over a forward-facing step, and  hypersonic flow over a circular cylinder, are carried out to demonstrate the performance of the proposed scheme.
\end{abstract}


\end{frontmatter}
\section{Introduction}
\label{Introduction}

In computational fluid dynamics (CFD) community, the gas kinetic scheme (GKS) based on the Boltzmann BGK model \cite{Bhatnagar1954A} for accurate solution in compressible and incompressible flows developed by Xu (2001) \cite{Xu2001A} has been verified to be an accurate and robust method \cite{xiong2011numerical,yuan2015immersed,pan2016gas,li2017implementation}. Different from the schemes based on the Navier-Stokes equation, the gas distribution function in the GKS solver contains both equilibrium and non-equilibrium properties. To model particle-related nature, the transport and collision terms in the BGK model are coupled in description of the gas distribution function at a cell interface. For a more delicate description of a flow field with fewer mesh cells, there are several works about third-order GKS. Li et al. (2010) proposed the way for reconstruction of conservation variables using high-order polynomials \cite{Li2010A}. This method is validated under unstructured mesh and proved to be an accurate multi-dimensional scheme. Ren et al (2015) proposed a third-order multidimensional gas-kinetic method for viscous flow simulation based on discontinuous Galerkin scheme \cite{Ren2015A}. In their work, cases in two and three dimensions are accurately simulated. This method can obtain good results in high Mach number compressible viscous and heat conducting flows and the low speed high Reynolds number laminar flows. The methods mentioned above are one-stage gas evolution model. Here, we are going to construct a two-stage fourth-order scheme. In construction of schemes of higher order such as fourth-order methods, one-stage evolution model will complicate the formulation of GKS flux \cite{Tang2014A}. As an alternative, the Lax-Wendroff (L-W) method \cite{Lax1960Systems} is applied in time evolution. Pan et al (2016) proposed a two-stage fourth order gas-kinetic scheme \cite{Pan2016An}. Under structured grid, the weighted essentially nonoscillatory (WENO) scheme is used in reconstruction and evolution step, which is based on L-W method. They gave very good results in shock capturing and robustness of their scheme is also very well.

In structured grid, grid lines have to extend across the entire computation field because of block-structure. So many cells are placed in regions where they are not needed. Under this condition, it is hard to improve the computational efficiency. On the other hand, the cells in an unstructured mesh can cover any region according to required resolution \cite{Ebeida2006Fast}. The finite volume method (FVM) is also very flexible in treating manifold hybrid mesh. Many works have been done under unstructured hybrid mesh \cite{Pan2016A, Liu2016A}.

Many works on high order reconstruction for the Navier-Stokes equations have been done under unstructured mesh. Liu et al (2017) proposed a high-order distance weighted biased averaging procedure under unstructured mesh \cite{Liu2017Accuracy}. Accurate results are obtained in resolving discontinuity in flow fields. The scheme also has good convergence property. L. Gamet et al (1999) proposed a compact scheme on non-uniform meshes by a full inclusion of metrics in the coefficients of the scheme \cite{Gamet1999Compact}. Based on this method, turbulent flows are directly numerically simulated. A. Burbeau et al (2002) proposed a high-order discontinuous Galerkin (DG) method under unstructured mesh \cite{Burbeau2002Simulation}. In their work, good accuracy and robustness are verified for subsonic and supersonic flows. Based on the cubic spline interpolation, Wang et al(2015) proposed a fourth-order scheme \cite{Wang2015An}.

In this work, the spline interpolation is applied in reconstruction for GKS on unstructured mesh. It combines properties of compact scheme and Hermite interpolation thus good spectral property and high-order accuracy are obtained. To achieve fourth order in space, the cubic spline interpolation introduced by Schoenberg \cite{schoenberg1988contributions} (1946) is used in reconstruction. It has first and second derivatives uniformly convergent to obtain sufficiently smooth functions. Compared to other ways of reconstruction, the cubic spline over the whole computation field can be obtained efficiently with a sparse matrix system. In many cases, the accuracy of derivatives obtained at the boundary by this method is very high \cite{Rubin1975Higher}. Under the orthogonal grid, the first derivative and the flux can meet the requirement of four order accuracy.

This paper is organized as follows. In Section \ref{sec:HGKS}, the GKS and its two stage temporal discretization are presented. In Section \ref{sec:SPLINE}, the cubic spline reconstruction and its coupling with shock-capturing WENO scheme are introduced. In Section \ref{sec:NUMERICAL RESULTS}, several test cases are conducted to validate the present scheme. The conclusion will be grouped in the last section.

\section{TWO STAGE GAS-KINETIC SCHEME}\label{sec:HGKS}
\subsection{GAS-KINETIC SCHEME}\label{sec:GKS}
The two-dimensional BGK equation can be written as~\cite{Xu2001A}
\begin{equation}\label{Eq01}
\frac{{\partial f}}{{\partial t}} + u\frac{{\partial f}}{{\partial x}} + v\frac{{\partial f}}{{\partial y}} = \frac{{g - f}}{\tau },
\end{equation}
Where  $f$  denotes the gas distribution function, $g$ denotes the equilibrium Maxwell distribution function, which are the functions of space $(x, y)$, particle velocity ${\vec u} = (u,v)$, time $t$, and internal variable $\xi$. $\tau$ is the relaxation time, which describes the average time between two times of particle collisions. The BGK model equation can be discretized in time and the gas distribution function at a cell interface $\left( x_{cf}, y_{cf} \right)$ can be written as
\begin{equation}\label{Eq02}
{f_{cf}}\left( {t, u, v, \xi } \right) = \frac{1}{\tau }\int_0^t {g\left( {x', y', u, v, \xi } \right){e^{ - \frac{{t - t'}}{\tau }}}dt'}  + {e^{ - \frac{t}{\tau }}}{f_0}\left( {{x_{cf}} - ut, {y_{cf}} - vt, \xi } \right),
\end{equation}
where subscript $cf$ denotes cell interface, $x' = {x_{cf}} - u\left( {t - t'} \right)$, $y' = {y_{cf}} - v\left( {t - t'} \right)$ denote particle trajectory. In above equations, the Maxwell distribution function reads
\begin{equation}\label{Eq03}
g = \rho {\left( {\frac{\lambda }{\pi }} \right)^{\frac{{K + 2}}{2}}}{e^{ - \lambda \left( {{{\left( {u - U} \right)}^2} + {{\left( {v - V} \right)}^2} + {\xi ^2}} \right)}},
\end{equation}
where $K$ is internal freedom degree. For diatomic gas, $K = 3$. $\lambda $ denotes internal energy. It has relation with $\xi $
\begin{equation}\label{Eq04}
\int {{\xi ^2}gdudvd\xi }  = \frac{{K + 2}}{{2\lambda }}.
\end{equation}
The conservation variables can be obtain via
\begin{equation}\label{Eq05}
{\bm W} = \left( \begin{array}{c}
\rho \\
\rho U\\
\rho V\\
\rho E
\end{array} \right) = \int {{\bm \psi} gdudvd\xi },
\end{equation}
where the macroscopic variables $\bm W = {\left( {\rho , \rho U, \rho V, \rho E} \right)^T}$, $\rho E = \frac{1}{2}\rho \left( {{U^2} + {V^2} + \frac{{K + 2}}{{2\lambda }}} \right)$, and ${\bm \psi} =(1, u, v, 1/2(u^2+v^2+\xi ^2))^T$. The pressure and density can be related with $\lambda $ as  $p = {\rho  \mathord{\left/ {\vphantom {\rho  {2\lambda }}} \right. \kern-\nulldelimiterspace} {2\lambda }}$.

The gas distribution function at the beginning of a time step $f_0$ can be assumed to have two spatial slopes on two sides of a cell interface \cite{Pan2016An}. To describe the real physical situation inside shock wave. The deviation from an equilibrium distribution is used in this paper. So the initial gas distribution function $f_0$ is written as
\begin{equation}\label{Eq06}
{f_0} = \left\{ \begin{array}{l}
{g^l}\left( {1 + {a^l}\left( {{x_{cf}} - ut} \right) + {b^l}\left( {{y_{cf}} - vt} \right) - \tau \left( {{a^l}u + {b^l}v + {A^l}} \right)} \right),    x \le {x_{cf}},\\
{g^r}\left( {1 + {a^r}\left( {{x_{cf}} - ut} \right) + {b^r}\left( {{y_{cf}} - vt} \right) - \tau \left( {{a^r}u + {b^r}v + {A^r}} \right)} \right),    x > {x_{cf}},
\end{array} \right.
\end{equation}
where $l$ and $r$ denote the left and right of a cell interface, respectively. Derivative terms $a$ and $b$ are spatial derivative of a Maxwellian. $A$ is its temporal derivatives. They have a unique correspondence with the slopes of the conservative variables,
\begin{equation}\label{Eq07}
\begin{aligned}
\int {a{\bm \psi} gdudvd\xi } & = \frac{{\partial \bm W}}{{\partial x}} ,\\
\int {b{\bm \psi} gdudvd\xi } & = \frac{{\partial \bm W}}{{\partial y}} ,\\
\int {A{\bm \psi} gdudvd\xi } & = \frac{{\partial \bm W}}{{\partial t}}.
\end{aligned}
\end{equation}

The derivative terms in Eq.~(\ref{Eq07}) can be written as Taylor expansion of a Maxwellian about particle velocities and has the following form
\begin{equation}\label{Eq08}
\begin{aligned}
a &= {a_0} + {a_1}u + {a_2}v + {a_3}\frac{1}{2}\left( {{u^2} + {v^2} + {\xi ^2}} \right) ,\\
b &= {b_0} + {b_1}u + {b_2}v + {b_3}\frac{1}{2}\left( {{u^2} + {v^2} + {\xi ^2}} \right) ,\\
A &= {A_0} + {A_1}u + {A_2}v + {A_3}\frac{1}{2}\left( {{u^2} + {v^2} + {\xi ^2}} \right).
\end{aligned}
\end{equation}
As for temporal derivatives ${A^l}$ and ${A^r}$, since the non-equilibrium parts have no direct contribution to the conservative variables, so we have
\begin{equation}\label{Eq09}
\begin{aligned}
\int {\left( {{a^l}u + {b^l}v + {A^l}} \right){\bm \psi} {g^l}dudvd\xi  = 0} ,\\
\int {\left( {{a^r}u + {b^r}v + {A^r}} \right){\bm \psi} {g^r}dudvd\xi  = 0} .
\end{aligned}
\end{equation}
All the Maxwellians in above equations should be decided via conservation variables ${\bm W}^l$ and ${\bm W}^r$ on both sides of a cell interface. They all are obtained by reconstruction. In this paper, the cubic spline interpolation is applied, which will be discussed in Section \ref{sec:SPLINE}.

The equilibrium distribution function in Eq.~(\ref{Eq02}) also has two slopes on both sides of a cell interface,
\begin{equation}\label{Eq10}
g = {g_0}\left( {1 + \left( {1 - H\left[ x \right]} \right){{\bar a}^l}x + H\left[ x \right]{{\bar a}^r}x + \bar by + \bar At} \right),
\end{equation}
where $H\left[ x \right]$ is Heaviside function which reads
\begin{equation}\label{Eq11}
H\left[ x \right] = \left\{ \begin{array}{l}
0, x < 0,\\
1,  x \ge 0 .
\end{array} \right.
\end{equation}
The spatial derivatives in Eq.~(\ref{Eq10}) are computed by
\begin{equation}\label{Eq12}
\begin{aligned}
\int {{{\bar a}^l}{\bm \psi} {g_0}dudvd\xi }  = \frac{{{{\bm W}_0} - {\bm W}_c^l}}{{{x_{cf}} - x_c^l}} ,\\
\int {{{\bar a}^r}{\bm \psi} {g_0}dudvd\xi }  = \frac{{{\bm W}_c^r - {{\bm W}_0}}}{{x_c^r - {x_{cf}}}}.
\end{aligned}
\end{equation}
And
\begin{equation}\label{Eq13}
\int {\bar b{g_0}dudvd\xi }  = \int_{u \ge 0} {{b^l}{g^l}} dudvd\xi  + \int_{u < 0} {{b^r}{g^r}} dudvd\xi ,
\end{equation}
where ${\bm W}_c^l$ and ${\bm W}_c^r$ are conservation variables of cells on the left and right of a cell interface.

Conservation variables ${\bm W}_0$ are given by
\begin{equation}\label{Eq14}
{\bm W}_0 = \int_{u \ge 0} {{g^l}{\bm \psi} dudvd\xi }  + \int_{u < 0} {{g^r}{\bm \psi} dudvd\xi } .
\end{equation}
Equilibrium state $g_0$ is computed via ${\bm W}_0$.

The term $\bar A$ is decided by constraint condition
\begin{equation}\label{Eq15}
\int_0^{\Delta t} {\int {\left( {g - {f_{cf}}} \right){\bm \psi} dudvd\xi } dt}  = 0 .
\end{equation}

After all unknown terms are obtained, the gas distribution function at a cell interface is written as
\begin{equation}\label{Eq16}
\begin{aligned}
{f_{cf}}\left( {t, u, v, \xi } \right) = &\left( {1 - {e^{ - \frac{t}{\tau }}}} \right){g_0} + \left( {\tau \left( { - 1 + {e^{ - \frac{t}{\tau }}}} \right) + t{e^{ - \frac{t}{\tau }}}} \right)\left( {{{\bar a}^l}H\left[ u \right] + {{\bar a}^r}\left( {1 - H\left[ u \right]} \right)} \right)u{g_0}\\
                      &+  \left( {\tau \left( { - 1 + {e^{ - \frac{t}{\tau }}}} \right) + t{e^{ - \frac{t}{\tau }}}} \right)\bar bv{g_0}\\
                      &+ \tau \left( {\frac{t}{\tau } - 1 + {e^{ - \frac{t}{\tau }}}} \right)\bar A{g_0}\\
                      &+ {e^{ - \frac{t}{\tau }}}\left( {1 - u\left( {t + \tau } \right){a^l} - v\left( {t + \tau } \right){b^l}} \right)H\left[ u \right]{g^l}\\
                      &+ {e^{ - \frac{t}{\tau }}}\left( {1 - u\left( {t + \tau } \right){a^r} - v\left( {t + \tau } \right){b^r}} \right)\left( {1 - H\left[ u \right]} \right){g^r}\\
                      &+ {e^{ - \frac{t}{\tau }}}\left( { - \tau {A^l}H\left[ u \right]{g^l} - \tau {A^r}\left( {1 - H\left[ u \right]} \right){g^r}} \right).\\
\end{aligned}
\end{equation}

Now, the numerical flux can be obtained via Eq.~(\ref{Eq16}). Next, to update the flow field, a two stage method proposed by Pan et al (2016) \cite{Pan2016An} is used. In gas nature, the evolution process is a relaxation from kinetic to hydrodynamic scale. To obtain time derivative of flux, the flux function within a time interval $\left[ {{t_n}, {t_n} + \Delta t} \right]$ is written as
\begin{equation}\label{Eq17}
{\bm \Phi} \left( {{{\bm W}^n}, \Delta t} \right) = \int_{{t_n}}^{{t_n} + \Delta t} {u{\bm \psi} {f_{cf}}\left( {t, u, v, \xi } \right)dt}  = \int_{{t_n}}^{{t_n} + \Delta t} {{\bm F}\left( {{{\bm W}^n}, t} \right)dt}.
\end{equation}

In this time interval, the flux term ${\bm F}\left( {{{\bm W}^n}, t} \right)$ is expanded in time
\begin{equation}\label{Eq18}
{\bm F}\left( {{{\bm W}^n}, t} \right) = {\bm F}\left( {{{\bm W}^n}, {t_n}} \right) + \frac{{\partial {\bm F}\left( {{{\bm W}^n}, {t_n}} \right)}}{{\partial t}}\left( {t - {t_n}} \right).
\end{equation}

After substituting Eq.~(\ref{Eq18}) into Eq.~(\ref{Eq17}), the flux in the time intervals $\left[ {{t_n}, {t_n} + \frac{{\Delta t}}{2}} \right]$ and $\left[ {{t_n}, {t_n} + \Delta t} \right]$ are presented as
\begin{equation}\label{Eq19}
\begin{aligned}
{\bm \Phi} \left( {{{\bm W}^n}, \frac{{\Delta t}}{2}} \right) &= \frac{1}{2}{\bm F}\left( {{W^n}, {t_n}} \right)\Delta t + \frac{1}{8}\frac{{\partial {\bm F}\left( {{{\bm W}^n}, {t_n}} \right)}}{{\partial t}}\Delta {t^2},\\
{\bm \Phi} \left( {{{\bm W}^n}, \Delta t} \right) &= {\bm F}\left( {{{\bm W}^n}, {t_n}} \right)\Delta t + \frac{1}{2}\frac{{\partial {\bm F}\left( {{{\bm W}^n}, {t_n}} \right)}}{{\partial t}}\Delta {t^2}.
\end{aligned}
\end{equation}
Next, the conservation variables at $t_* = {t_n} + \frac{{\Delta t}}{2}$ are computed,
\begin{equation}\label{Eq20}
{{\bm W}^*} = {{\bm W}^n} - \frac{1}{\Omega }\int {{\bm \Phi}\left( {{{\bm W}^n}, \frac{{\Delta t}}{2}} \right) dS},
\end{equation}
where $\Omega $ is the volume of a cell and $S$ denotes the area of a interface.

Similarly, we have
\begin{equation}\nonumber
\begin{aligned}
{\bm \Phi} \left( {{{\bm W}^*}, \frac{{\Delta t}}{4}} \right) &= \frac{1}{4}F\left( {{{\bm W}^*}, {t_*}} \right)\Delta t + \frac{1}{{32}}\frac{{\partial {\bm F}\left( {{{\bm W}^*}, {t_*}} \right)}}{{\partial t}}\Delta {t^2},\\
{\bm \Phi} \left( {{{\bm W}^*}, \frac{{\Delta t}}{2}} \right) &= {\bm F}\left( {{{\bm W}^*}, {t_*}} \right)\frac{{\Delta t}}{2} + \frac{1}{8}\frac{{\partial {\bm F}\left( {{{\bm W}^*}, {t_*}} \right)}}{{\partial t}}\Delta {t^2}.
\end{aligned}
\end{equation}

Finally, the flow field is updated by
\begin{equation}\label{Eq21}
{{\bm W}^{n + 1}} = {{\bm W}^n} - \frac{{\Delta t}}{\Omega }\int {{\bm \phi}\left( {{{\bm W}^n}, {t_n}} \right) dS}.
\end{equation}
To construct temporal numerical flux $\bm \phi $, the Runge-Kutta scheme is used to compute its time derivative. The flux term at $t_n$ and $t_*$ are applied to obtain a two-stage advancing method.
\begin{equation}\label{Eq22}
{\bm \phi}\left( {{{\bm W}^n}, {t_n}} \right) = {\bm F}\left( {{{\bm W}^n}, {t_n}} \right) + \frac{{\Delta t}}{6}\left( {\frac{{\partial {\bm F}\left( {{{\bm W}^n}, {t_n}} \right)}}{{\partial t}} + 2\frac{{\partial {\bm F}\left( {{{\bm W}^*}, {t_*}} \right)}}{{\partial t}}} \right).
\end{equation}

\section{CUBIC SPLINE RECONSTRUCTION}\label{sec:SPLINE}
In Section \ref{sec:HGKS}, the equilibrium distribution function at a cell interface should be obtained from reconstruction of conservation variables. The cubic spline interpolation is used because it can give highly accurate results. In this paper, the reconstruction based on the cubic spline developed by Wang et al (2015) \cite{Wang2015An} is applied. First, we should have a fitting formula for two-dimensional cases
\begin{equation}\label{Eq23}
w\left( {x, y} \right) = {c_0} + {c_1}x + {c_2}y + {c_3}{x^2} + {c_4}{y^2} + {c_5}{x^3} + {c_6}{y^3}.
\end{equation}

To determine the value of unknown coefficients in Eq.~(\ref{Eq23}), the interpolation conditions are given \cite{Li2015A}. In flow fields, we have
\begin{equation}\label{Eq24}
\begin{aligned}
{\bm w}\left( {{x_0}, {y_0}} \right) &= {\bm W}\left( {{x_0}, {y_0}} \right),\\
{{{\bm w}'}^l} &= {{{\bm w}'}^r},\\
{{{\bm w}''}^l} &= {{{\bm w}''}^r}.
\end{aligned}
\end{equation}

On the boundaries in flow fields, different types of boundary conditions should be implemented \cite{Li2015A}:

1 Inflow boundary condition
\begin{equation}\label{Eq25}
\begin{aligned}
{\bm w}\left( {{x_0}, {y_0}} \right) &= {\bm W}\left( {{x_0}, {y_0}} \right),\\
{{\bm w}_{cf}} &= \frac{1}{{12}}\left( { - {\bm W}\left( {{x_1}, {y_1}} \right) + 7{\bm W}\left( {{x_0}, {y_0}} \right) + 7{\bm W}\left( {{x_{ - 1}}, {y_{ - 1}}} \right) - {\bm W}\left( {{x_{ - 2}}, {y_{ - 2}}} \right)} \right),\\
{{{\bm w}'}_{cf}} &= \frac{1}{{12\Delta x}}\left( { - {\bm W}\left( {{x_1}, {y_1}} \right) + 15{\bm W}\left( {{x_0}, {y_0}} \right) - 15{\bm W}\left( {{x_{ - 1}}, {y_{ - 1}}} \right) + {\bm W}\left( {{x_{ - 2}}, {y_{ - 2}}} \right)} \right).
\end{aligned}
\end{equation}

2 Outflow boundary condition
\begin{equation}\label{Eq26}
\begin{aligned}
{\bm w}\left( {{x_0}, {y_0}} \right) &= {\bm W}\left( {{x_0}, {y_0}} \right),\\
{{\bm w}_{cf}} &= \frac{1}{{12}}\left( {25{\bm W}\left( {{x_1}, {y_1}} \right) - 23{\bm W}\left( {{x_0}, {y_0}} \right) + 13{\bm W}\left( {{x_{ - 1}}, {y_{ - 1}}} \right) - 3{\bm W}\left( {{x_{ - 2}}, {y_{ - 2}}} \right)} \right),\\
{{{\bm w}'}_{cf}} &= \frac{1}{{12\Delta x}}\left( {35{\bm W}\left( {{x_1}, {y_1}} \right) - 69{\bm W}\left( {{x_0}, {y_0}} \right) + 45{\bm W}\left( {{x_{ - 1}}, {y_{ - 1}}} \right) - 11{\bm W}\left( {{x_{ - 2}}, {y_{ - 2}}} \right)} \right).
\end{aligned}
\end{equation}

3 Solid-wall boundary condition
\begin{equation}\label{Eq27}
\begin{aligned}
{\bm w}\left( {{x_0}, {y_0}} \right) &= {\bm W}\left( {{x_0}, {y_0}} \right),\\
{{\bm w}_{cf}} &=  - \frac{1}{{12}}\left( {25{\bm W}\left( {{x_1}, {y_1}} \right) - 23{\bm W}\left( {{x_0}, {y_0}} \right) + 13{\bm W}\left( {{x_{ - 1}}, {y_{ - 1}}} \right) - 3{\bm W}\left( {{x_{ - 2}}, {y_{ - 2}}} \right)} \right),\\
{{{\bm w}'}_{cf}} &=  - \frac{1}{{12\Delta x}}\left( {35{\bm W}\left( {{x_1}, {y_1}} \right) - 69{\bm W}\left( {{x_0}, {y_0}} \right) + 45{\bm W}\left( {{x_{ - 1}}, {y_{ - 1}}} \right) - 11{\bm W}\left( {{x_{ - 2}}, {y_{ - 2}}} \right)} \right).
\end{aligned}
\end{equation}

A sparse linear system of equations is constructed for whole flow field. After solving the system of equations, all the spatial derivatives for each cells are obtained, and the fitting formula in Eq.~(\ref{Eq23}) can be used for reconstruction.

Spurious oscillations usually occur near the discontinuity in the process of calculation. So the hybrid scheme in which cubic spline combined with WENO is applied to deal with the discontinuity in flow fields \cite{Ren2003A}. In the final system of equations for unknown coefficients in fitting formula, we have
\begin{equation}\label{Eq28}
{\bm R}' = \sigma {\bm R} + \left( {1 - \sigma } \right){{\bm m}^{WENO}},
\end{equation}
where ${\bm R}$ is the right hand side of this equation, $\sigma $ is the minmod function \cite{Shu1989Efficient}, and
\begin{equation}\label{Eq29}
{{\bm m}^{WENO}} = \frac{{1 + sign\left( {{U_n}} \right)}}{2}{\bm W}_{cf}^{WENO} + \frac{{1 - sign\left( {{U_n}} \right)}}{2}{\bm W}_{cf}^{WENO}.
\end{equation}
In Eq.~(\ref{Eq29}), ${\bm W}_{cf}^{WENO}$ is variable reconstructed by WENO scheme, ${U_n}$ is the velocity normal to the cell interface.

\section{NUMERICAL RESULTS}\label{sec:NUMERICAL RESULTS}
\subsection{Couette flow}\label{sec:Couette}
\begin{figure}
  \centering
  \includegraphics[width=0.5\textwidth]{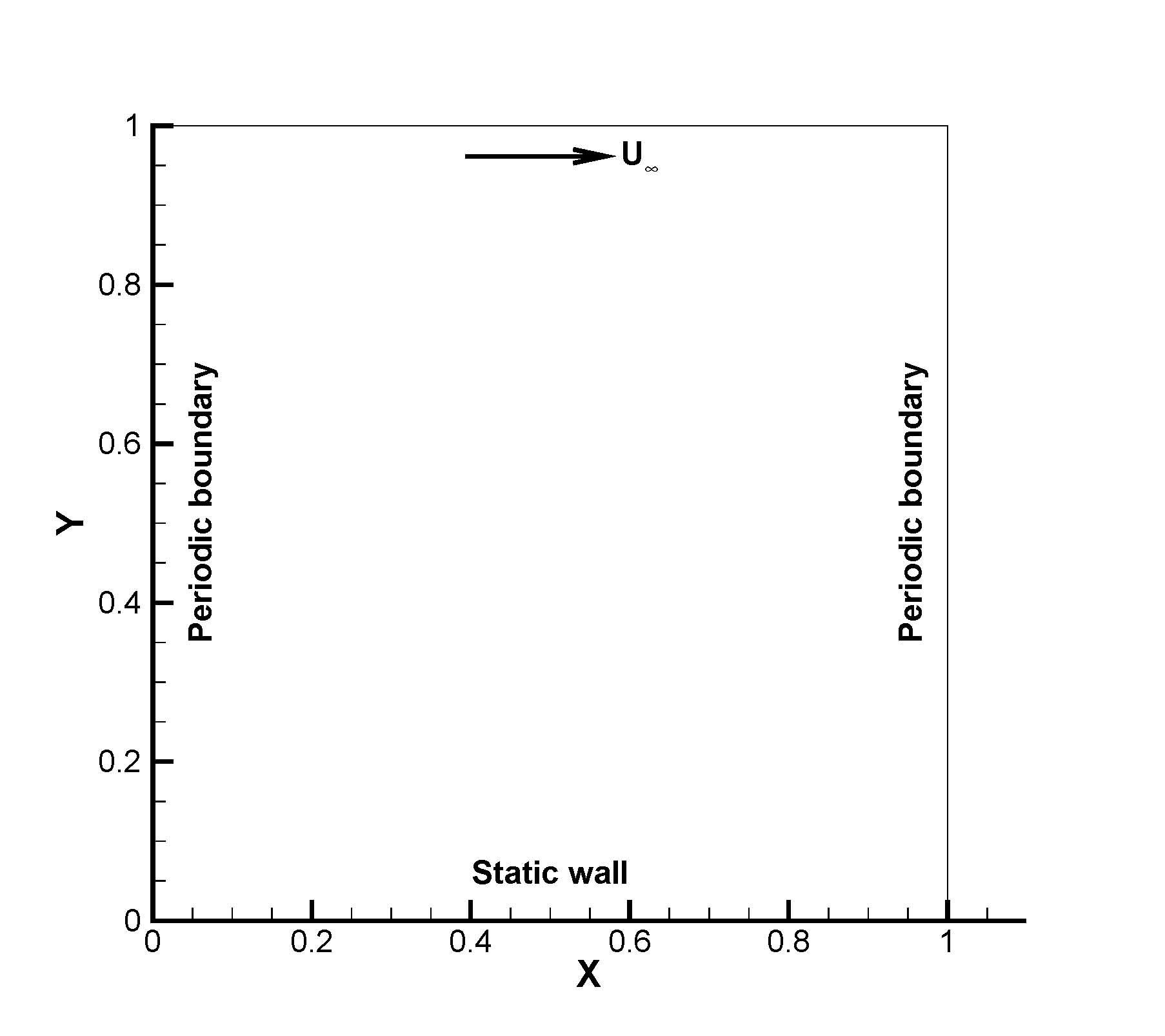}
  \caption{The computational domain for the Couette flow.}
  \label{fig:Fig01}
\end{figure}

For planar Couette flow, the fluid flows between a surface sliding with a constant velocity, ${U_\infty }$, and a stationary surface, is simulated under different mesh sizes to verify two-stage fourth-order GKS. The computational domain and its boundary conditions are shown in Fig.~\ref{fig:Fig01}.

\begin{figure}
  \centering
  \includegraphics[width=0.6\textwidth]{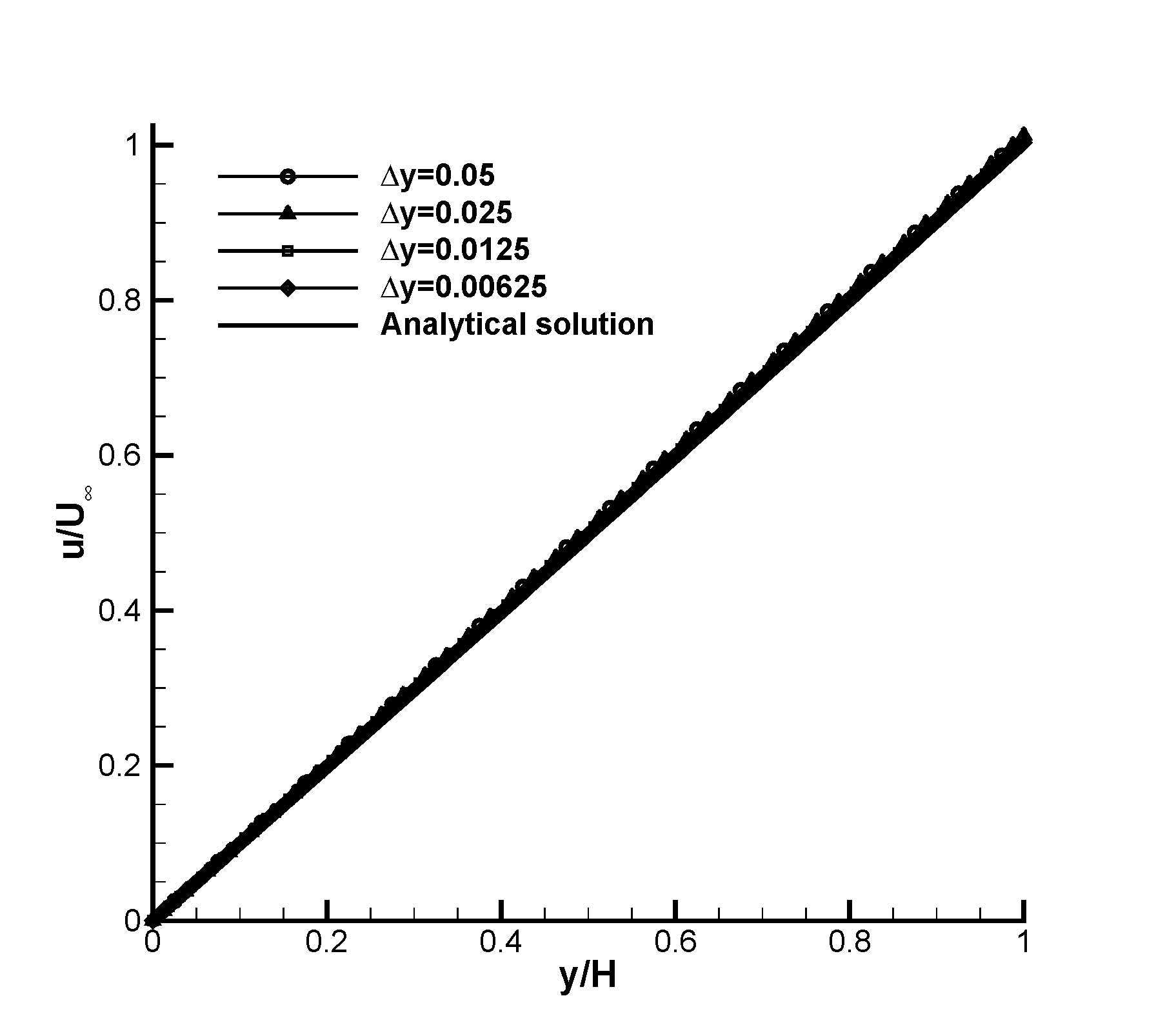}
  \caption{The comparison results for velocity profiles of the Couette flow.}
  \label{fig:Fig02}
\end{figure}

The reference length of Reynolds number is define as the height of the channel. Between two plates, 20, 40, 80, 160 cells are used for accuracy test. Velocity profiles under different grids and comparison results are given in Fig.~\ref{fig:Fig02}.

\begin{figure}
  \centering
  \includegraphics[width=0.5\textwidth]{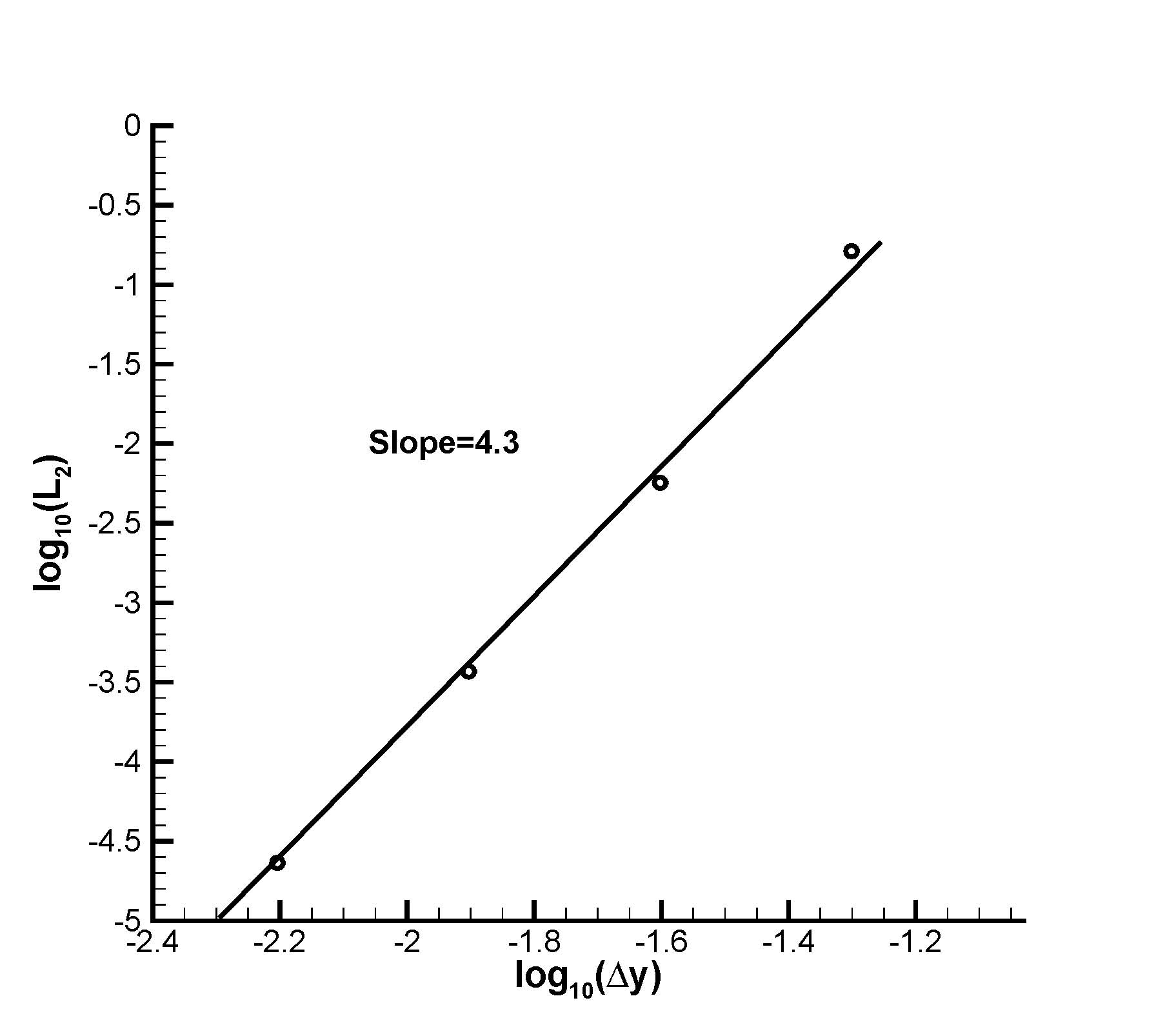}
  \caption{The L2 norm for residual of the Couette flow.}
  \label{fig:Fig03}
\end{figure}

L2 norm of residuals between the analytical solution and the numerical approaches under different grids is given in Fig.~\ref{fig:Fig03}. The slope of L2 norm varying with mesh size is about 4, which shows that the proposed scheme has fourth-order accuracy in space.

\subsection{Shu-Osher shock-acoustic wave interaction problem}\label{sec:Shu-Osher}

\begin{figure}
  \centering
  \includegraphics[width=0.5\textwidth]{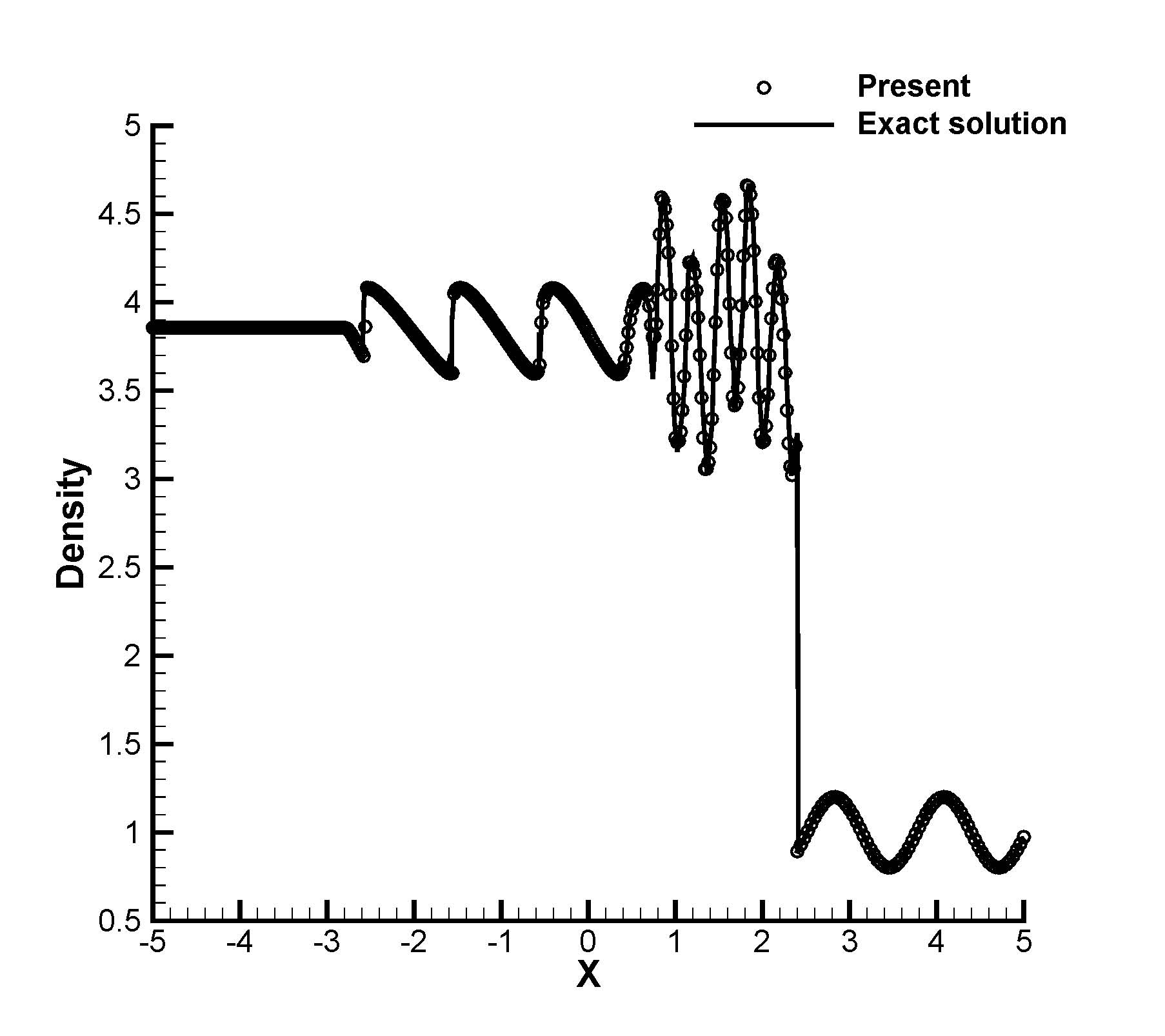}
  \caption{The comparison results for density profile of Shu-Osher problem.}
  \label{fig:Fig04}
\end{figure}

In this one-dimensional case, the computational domain in x-direction is $\left[ { - 5, 5} \right]$ and the initial flow field is given as \cite{Shu1989Efficient}
\begin{equation}\label{Eq30}
\left( {\rho , U, p} \right) = \left\{ \begin{array}{l}
\left( {3.857134, 2.629369, 10.33333} \right),    x \le  - 4,\\
\left( {1 + 0.2\sin \left( {5x} \right), 0, 1} \right),                        - 4 < x.
\end{array} \right.
\end{equation}
Comparison result of density profile at $t=1.8$ is shown in Fig.~\ref{fig:Fig04}. The grid size is 0.025 and the time step is ${10^{ - 4}}$.

This Shu-Osher's problem shows the difficulty of capturing both small-scale structure and shocks in flow. In previous Navier-Stokes solver, the artificial dissipation usually deteriorates description of delicate structure. To remove spurious dissipation, the GKS couples the transport and collision in resolving flow evolution. High-order reconstruction ensures exactness in modeling nonlinear phenomenon in flow structure.

\subsection{Woodward-Colella blast wave problem}\label{sec:Woodward-Colella blast}

Another one-dimensional Riemann problem is Woodward-Colella blast wave \cite{Jiang1996Efficient}, which consists of two discontinuities in pressure that lead to strong shock waves which interact with each other. The computational domain is $\left[ {0, 100} \right]$ which discretized with 800 cells. On both ends of the computational domain, the reflecting boundary condition is imposed. The initial condition is
\begin{equation}\label{Eq31}
\left( {\rho , U, p} \right) = \left\{ \begin{array}{l}
\left( {1, 0, 1000} \right),0 \le x < 10,\\
\left( {1, 0, 0.01} \right),10 \le x < 90,\\
\left( {1, 0, 100} \right),  90 \le x \le 100.
\end{array} \right.
\end{equation}

\begin{figure}
  \centering
  \includegraphics[width=0.5\textwidth]{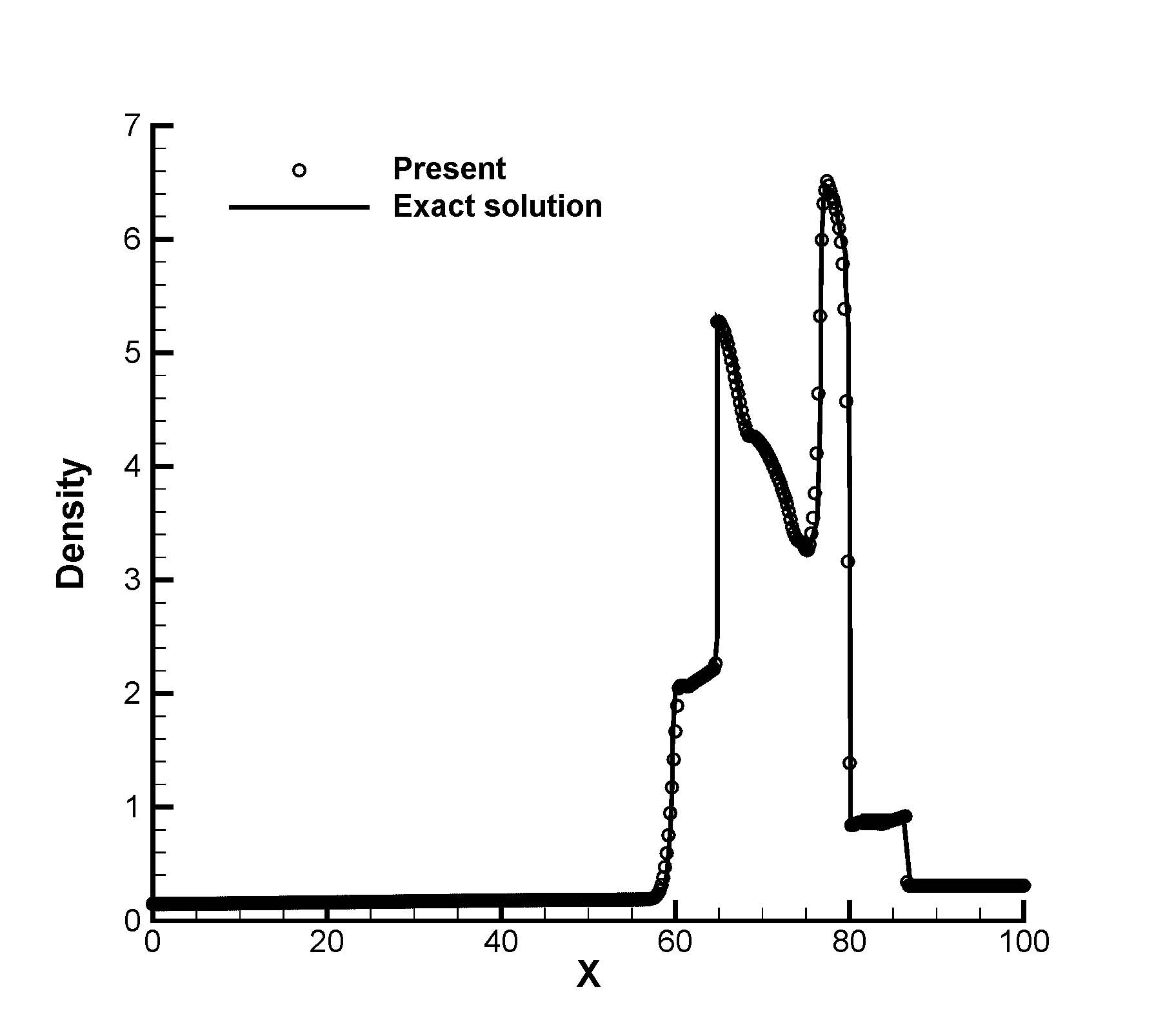}
  \caption{The comparison result for density profile of the Woodward-Colella blast problem.}
  \label{fig:Fig05}
\end{figure}

\begin{figure}
  \centering
  \includegraphics[width=0.5\textwidth]{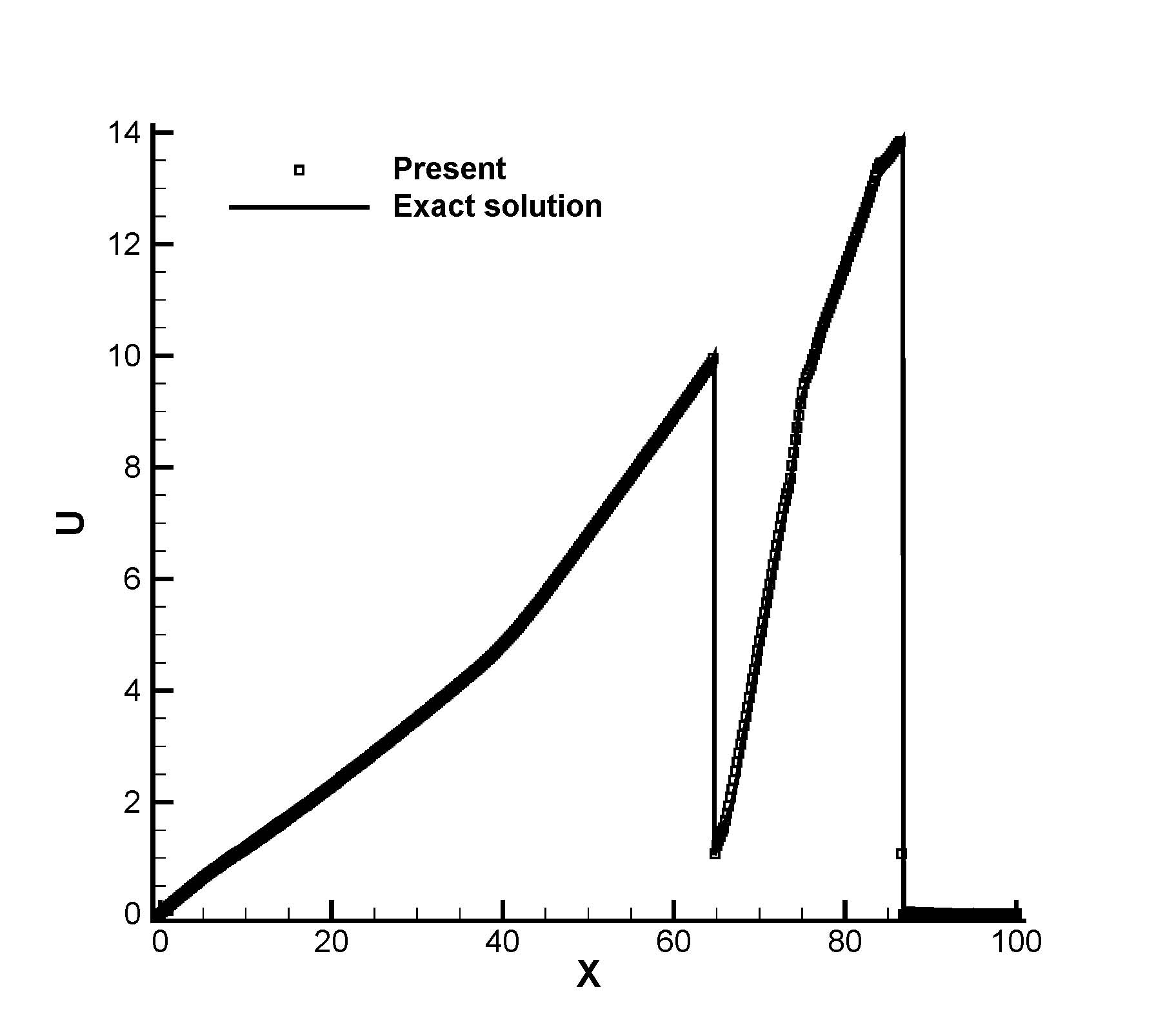}
  \caption{The comparison results for velocity profile of the Woodward-Colella blast problem.}
  \label{fig:Fig06}
\end{figure}

\begin{figure}
  \centering
  \includegraphics[width=0.5\textwidth]{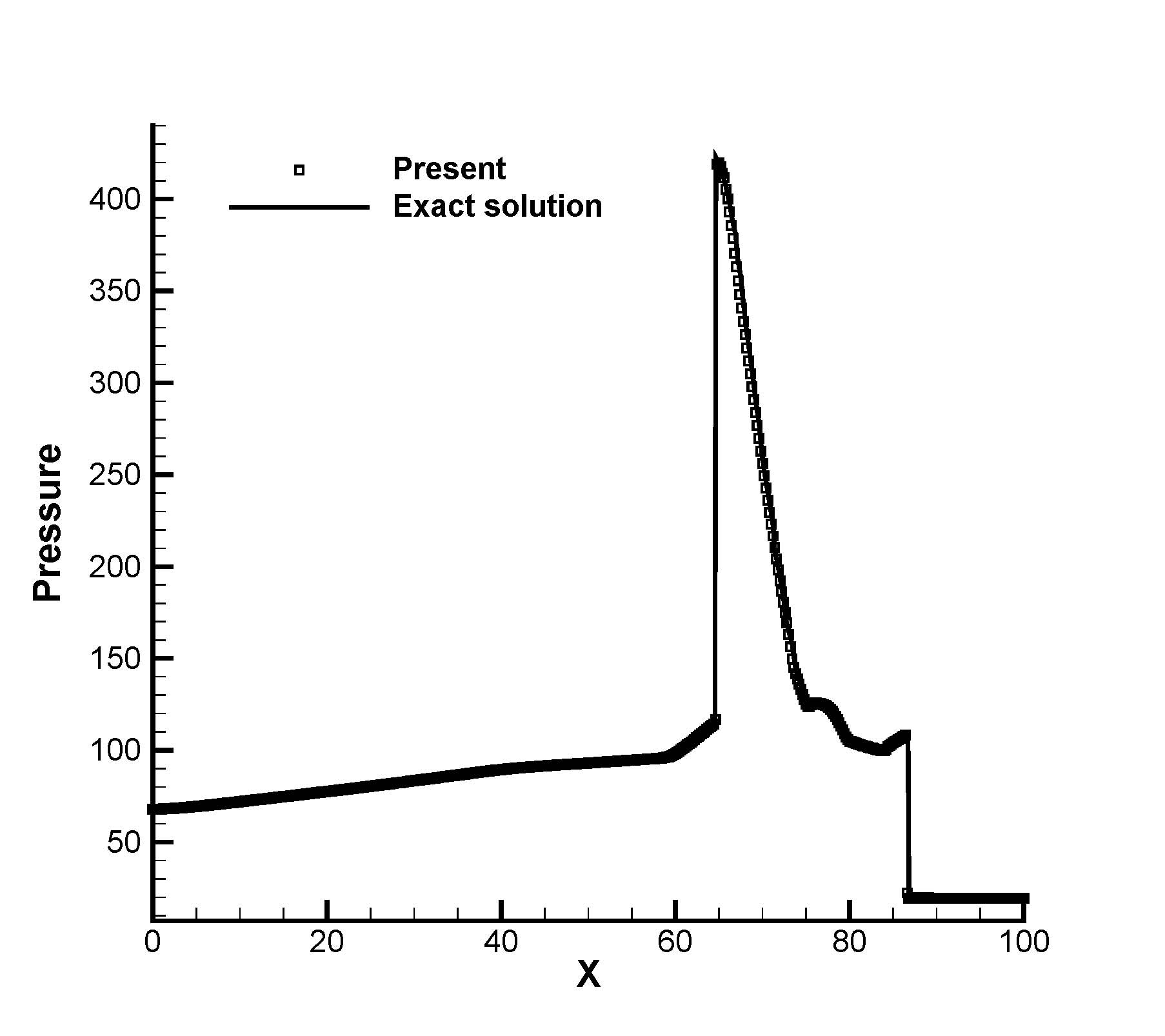}
  \caption{The comparison results for pressure profile of the Woodward-Colella blast problem.}
  \label{fig:Fig07}
\end{figure}

The density in initial flow field stay constant in space but there is huge discontinuity in pressure. So this benchmark case can be used to test both accuracy and robustness of the present scheme. The comparison results at $t=3.8$ with exact solution for density, velocity and pressure are given in Fig.~\ref{fig:Fig05}, Fig.~\ref{fig:Fig06} and Fig.~\ref{fig:Fig07} respectively.

Strong continuity with $10^5$ times of difference in pressure will challenge many Navier-Stokes solvers based on equilibrium assumption or flux vector splitting method. The deviation of equilibrium state leads to nonphysical results. Post-shock oscillation, shock wave instability, sonic point glitch and over-heating occur due to the above averaging dissipative mechanism in flux vector splitting method. In our work, shock waves are accurately and robustly resolved. Dynamical and kinematical dissipation are coupled in resolving flow discontinuities. During reconstruction, the high-order interpolation achieves high resolution in flux evaluation at the cell interface.

The cell size in uniform grid is 0.125 and the time step is ${10^{ - 5}}$. The results exhibit that the fourth-order GKS has a very good accuracy in shock capturing and good robust stability. In this case, the discontinuity is exactly captured with low dissipation and without oscillation. In next case, this property will be shown more notable.

\subsection{Two-dimensional Riemann problem}\label{sec:Two-dimensional Riemann problem}

\begin{figure}
  \centering
  \includegraphics[width=0.5\textwidth]{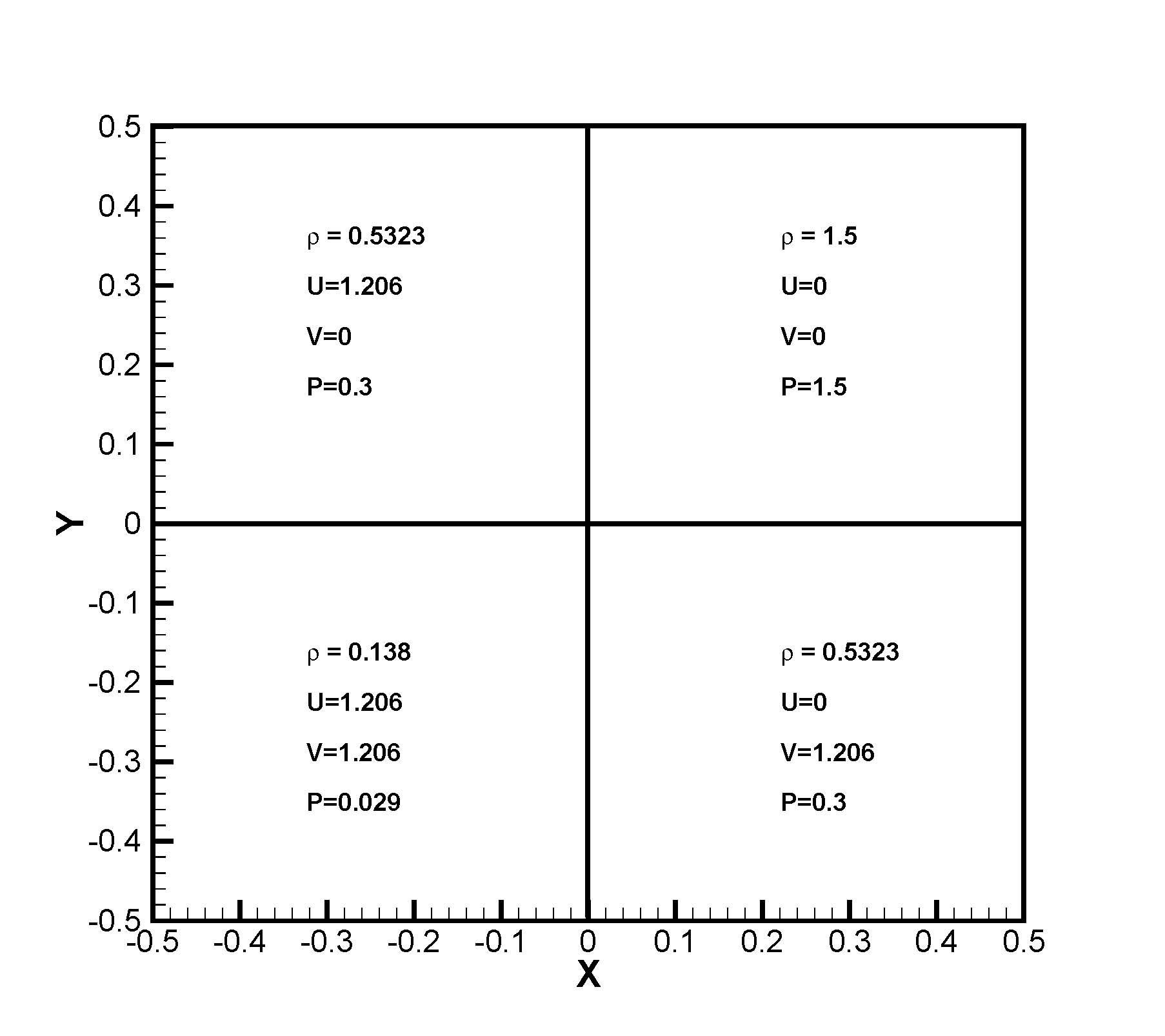}
  \caption{The initial condition of two-dimensional Riemann problem.}
  \label{fig:Fig08}
\end{figure}

In this part, the two-dimensional Riemann problem is carried out, which involves interactions of shocks \cite{Alexander2002Solution}. In this case, the two-stage fourth-order GKS is applied to capture details in shock interaction. The computational domain discretized by $400 \times 400$ grid is $\left[ { - 0.5, 0.5} \right] \times \left[ { - 0.5, 0.5} \right]$ and non-reflecting condition is implemented at all outer boundaries. The initial condition for this case is shown in Fig.~\ref{fig:Fig08}.

\begin{figure}
  \centering
  \includegraphics[width=0.5\textwidth]{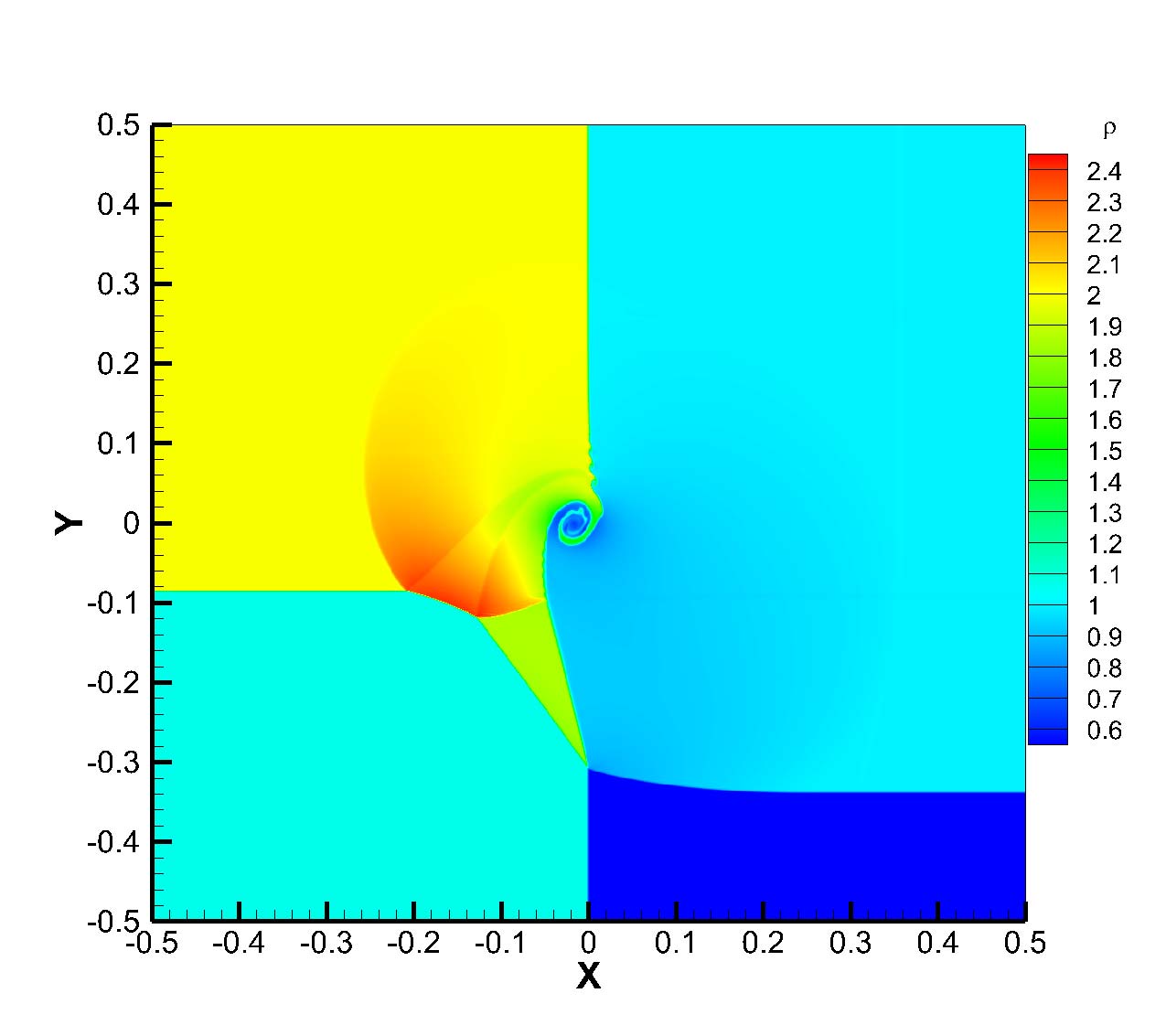}
  \caption{The density contour at t=0.3 in two-dimensional Riemann problem.}
  \label{fig:Fig09}
\end{figure}

\begin{figure}
  \centering
  \includegraphics[width=0.5\textwidth]{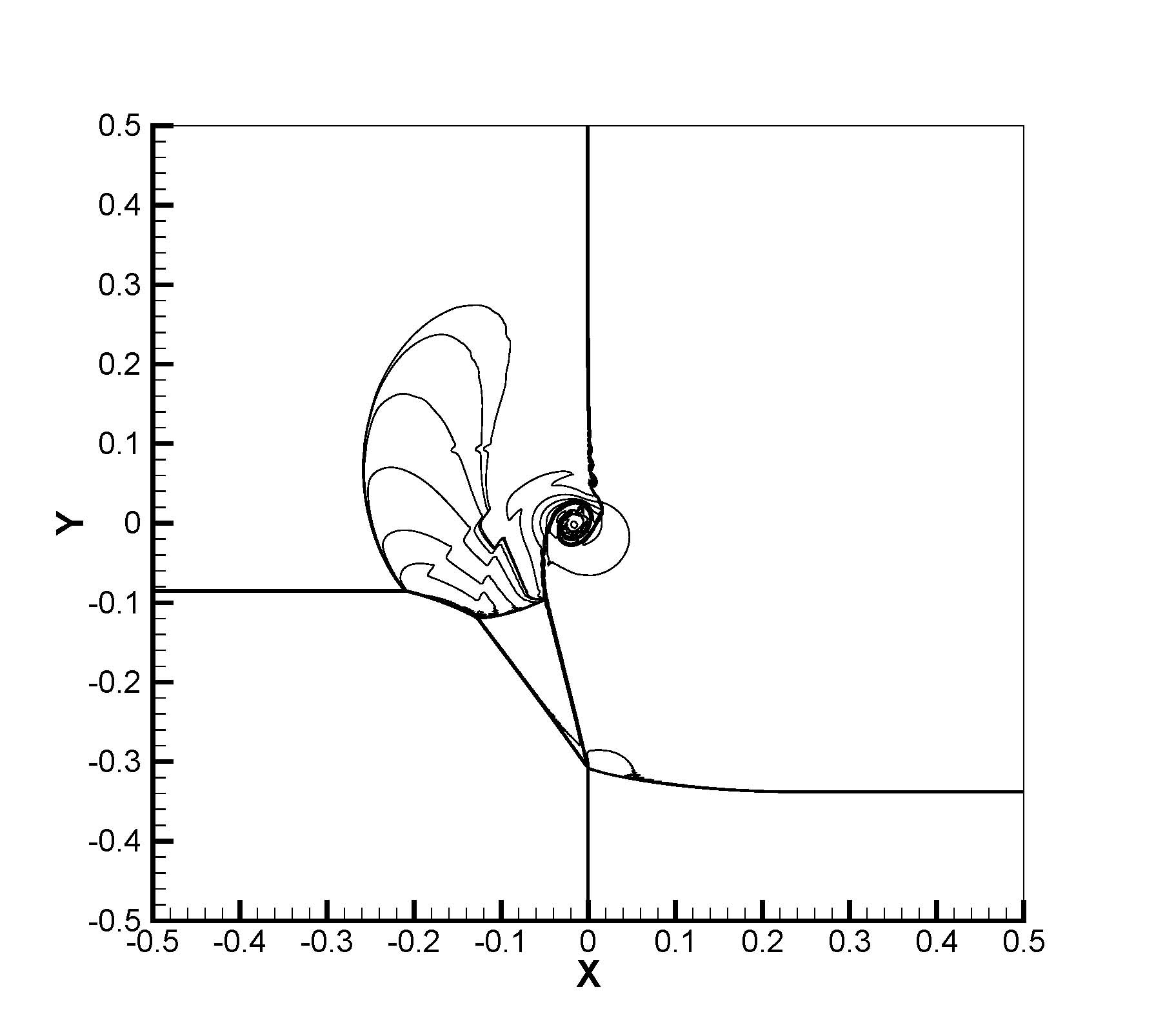}
  \caption{The shock interaction at t=0.3 in two-dimensional Riemann problem.}
  \label{fig:Fig10}
\end{figure}

The density contour at $t=0.3$ is shown in Fig.~\ref{fig:Fig09}. The shock interaction is shown in Fig.~\ref{fig:Fig10}. Many delicate details are presented. In traditional Navier-Stokes solver based on equilibrium state, it is still a difficult problem to exactly describe shock interaction and reflection without adding extra artificial dissipation. For GKS, the approach is keeping the deviation of Maxwellian and coupling particle transport and collision.

\subsection{Shock-boundary layer interaction}\label{sec:Shock-boundary layer interaction}

\begin{figure}
  \centering
  \includegraphics[width=0.5\textwidth]{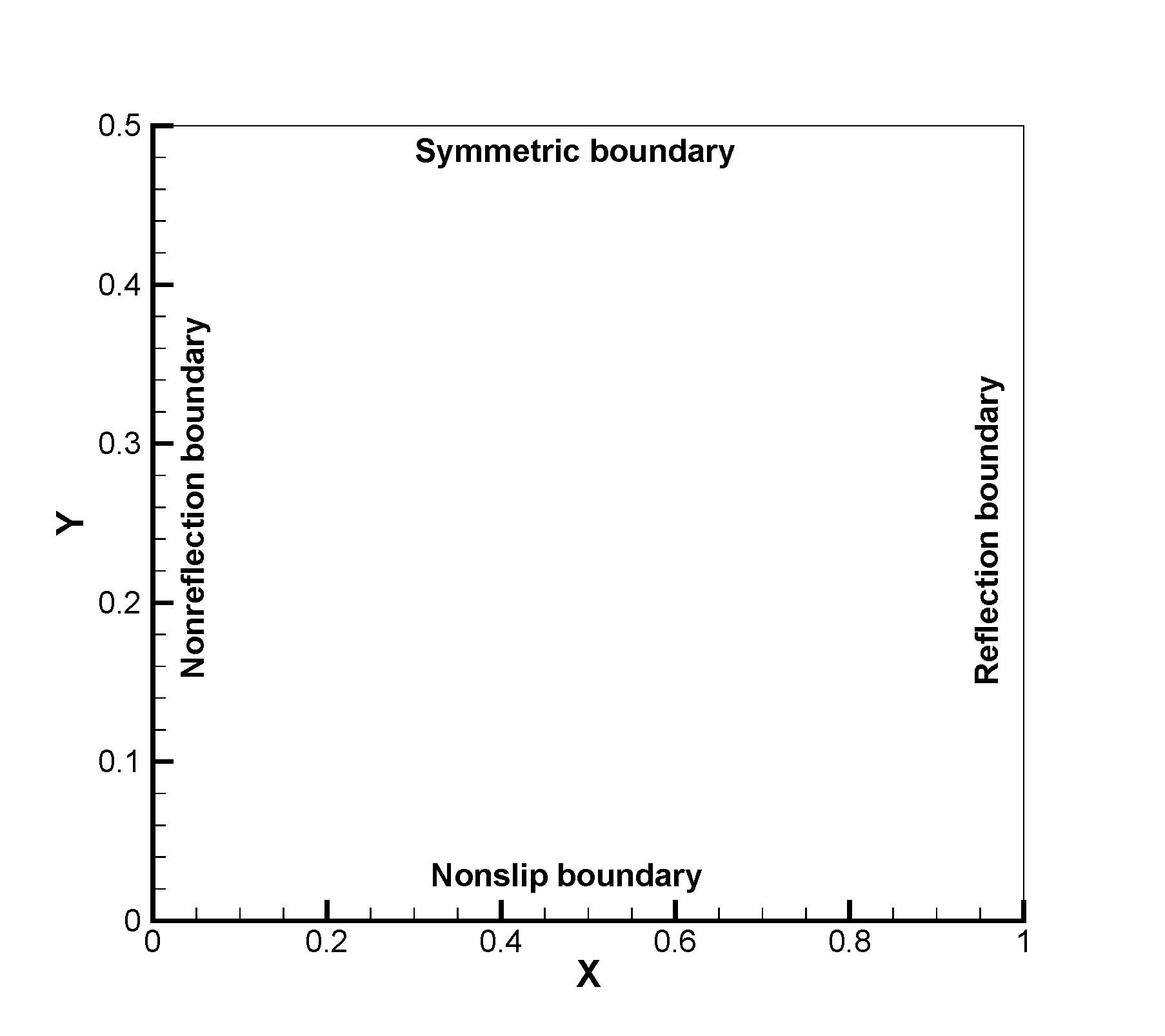}
  \caption{The computational domain for the viscous shock tube flow.}
  \label{fig:Fig11}
\end{figure}

This case is introduced to test the performance in simulation of shock-wave/boundary-layer interactions \cite{Daru2009Numerical}. Between two plates, a membrane is located at the middle of the channel. Two different states are separated by it. Interaction occurs when membrane is removed. The shock wave caused by contact discontinuity moves to the right and reflects at the end of the channel. The contact discontinuity and shock wave from reflection will interact with each other. The flow also causes a boundary layer near the wall. They will develop a shock-shock interaction and a shock-boundary layer interaction. The flow field is in a $\left[ {0, 1} \right] \times \left[ {0, 1} \right]$ box. In this paper, $\left[ {0, 1} \right] \times \left[ {0, 0.5} \right]$ is chose to be computational domain, and the symmetric boundary is implemented at the top, which is shown in Fig.~\ref{fig:Fig11}.

\begin{figure}
  \centering
  \subfigure[]{\label{fig:Fig12a}\includegraphics[width=0.45\textwidth]{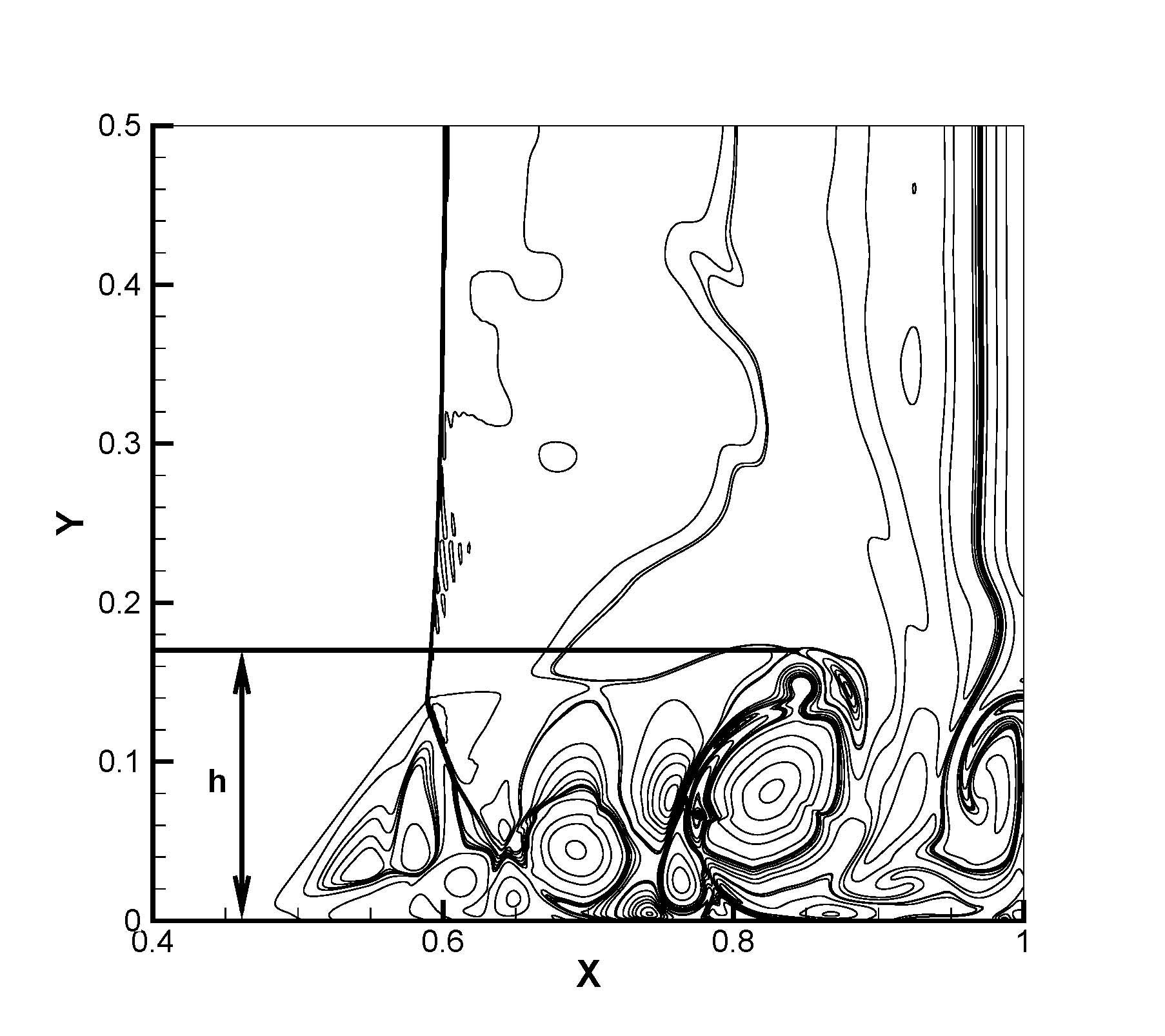}}
  \subfigure[]{\label{fig:Fig12b}\includegraphics[width=0.45\textwidth]{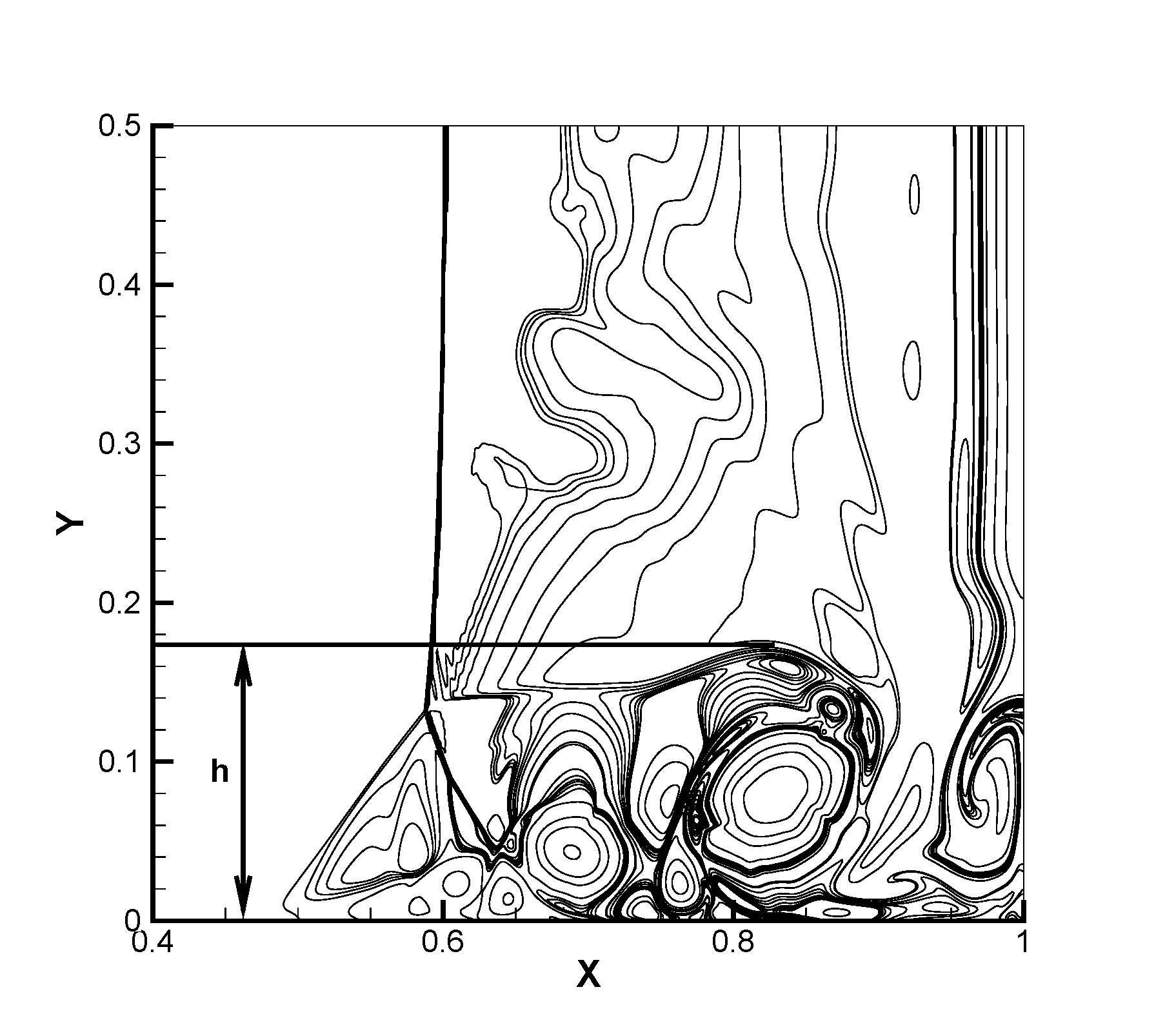}}
  \caption{The density distribution of the viscous shock tube flow. \\(a) $\Delta x= 1/500$, (b) $\Delta x= 1/750$.}
  \label{fig:Fig12}
\end{figure}

The initial condition is
\begin{equation}\label{Eq32}
\left( {\rho , U, p} \right) = \left\{ \begin{array}{l}
\left( {120, 0 ,120/\gamma } \right),     0 < x < 0.5,\\
\left( {1.2, 0, 1.2/\gamma } \right),       0.5 \le x < 1,
\end{array} \right.
\end{equation}
where $\gamma  = 1.4$, Mach number is 2.37 and the Reynolds number is 200. the size of the mesh cell is $\Delta x = 0.002$. The density distribution at $t=1$ is presented in Fig.~\ref{fig:Fig12a}. Under mesh in which cell size is $1/750$, the computational result is shown in Fig.~\ref{fig:Fig12b}.

The height of primary vortex agrees well with the results of Ref.~\cite{Kim2005Accurate}. On mesh in which cell size $\Delta x =1/500$, the height of primary vortex computed in Ref.~\cite{Kim2005Accurate} under advection upstream splitting method by pressure-based weight function (AUSMPW) is $0.163$. The result from the improved M-AUSMPW is better, which is $0.168$. Under the same mesh, the two-stage four-order WENO GKS \cite{Pan2016An} obtains a height of $0.171$. In this case, we also obtain $0.17$. Compared to fifth-order WENO scheme based on Navier-Stokes equations in Ref.~\cite{Daru2000Evaluation}, in which the height of primary vortex is $0.16$. the GKS also has less dissipation and captures more details in shock/shear/boundary-layer interactions. To show a more complicated flow structure, this case is tested under mesh in which cell size $\Delta x =1/750$. The height of primary vortex is $0.175$. This result is consistent with Ref.~\cite{Pan2016An}.

In this case, the numerical results indicate good shock transition without any noticeable oscillation. In this unsteady flow with strong discontinuities, the spatial and temporal precision should be strictly ensured to capture the gas evolution. The two-stage four-order GKS presents many physical details in shock interaction, which demonstrates that both the high-order initial reconstruction and high-order gas evolution model are important in CFD simulations.

\subsection{Forward-facing step flow}\label{sec:forward-facing step}

\begin{figure}
  \centering
  \includegraphics[width=0.5\textwidth]{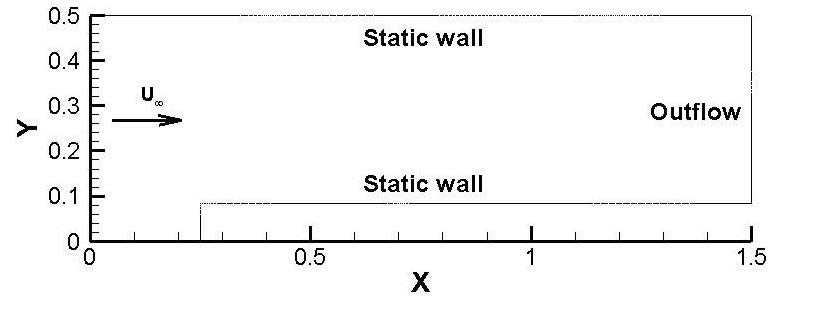}
  \caption{The boundary condition for the flow over a forward-facing step.}
  \label{fig:Fig13}
\end{figure}

In this case, unstructured hybrid mesh with adaptive mesh refinement (AMR) is applied for shock-capturing. The supersonic flow passes through a forward-facing step whose height is $1/6$ that of computational domain. The ratio of length to height is $3:1$. The boundary condition is shown in Fig.~\ref{fig:Fig13}.

\begin{figure}
  \centering
  \includegraphics[width=0.5\textwidth]{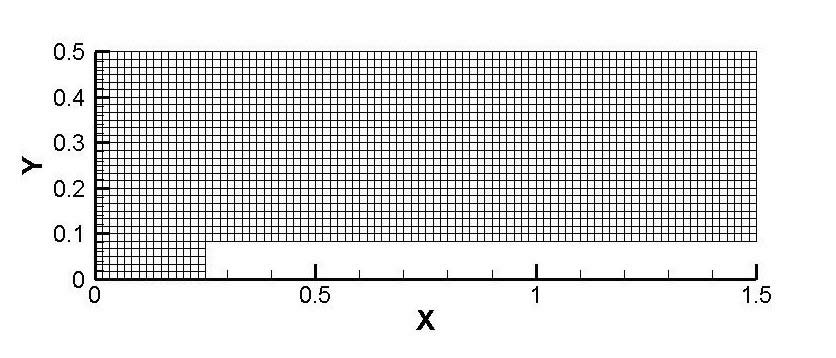}
  \caption{The initial mesh for the flow over a forward-facing step.}
  \label{fig:Fig14}
\end{figure}

\begin{figure}
  \centering
  \includegraphics[width=0.5\textwidth]{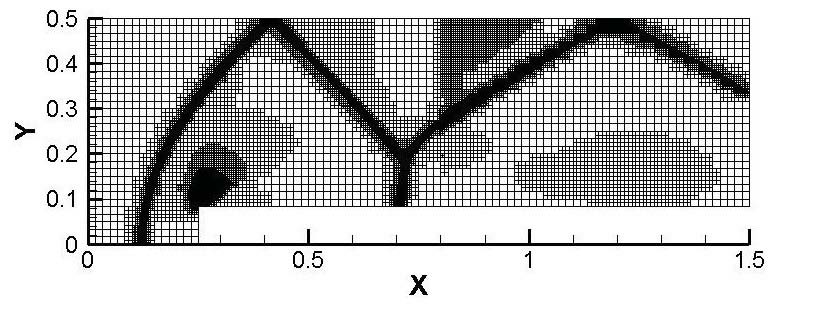}
  \caption{The refined mesh for the flow over a forward-facing step.}
  \label{fig:Fig15}
\end{figure}

The Mach number of free stream is 3.0 and the Reynolds number is 1000. The computation is started under a coarse mesh with 2325 cells. After convergence, the coarse mesh is refined within the region where gradient and vorticity are greater than mean level. After refinement, there are quadrilaterals, pentagons and hexagons in the mesh. The total number of cells is finally 43368. The initial mesh and refined hybrid mesh are given in Fig.~\ref{fig:Fig14} and Fig.~\ref{fig:Fig15} respectively.

\begin{figure}
  \centering
  \includegraphics[width=0.5\textwidth]{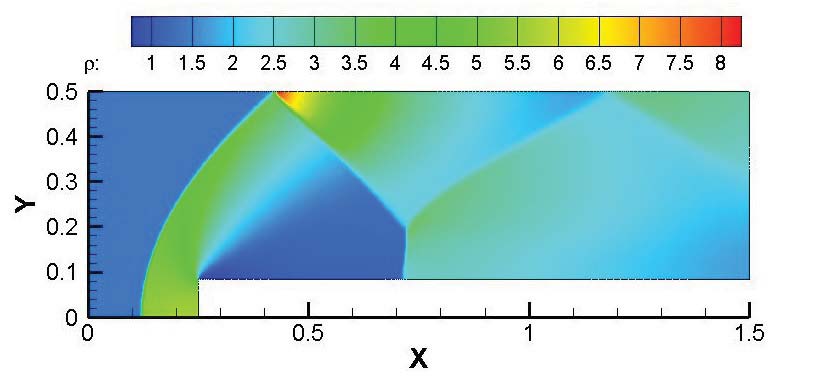}
  \caption{The density distribution of the flow over a forward-facing step.}
  \label{fig:Fig16}
\end{figure}

\begin{figure}
  \centering
  \includegraphics[width=0.5\textwidth]{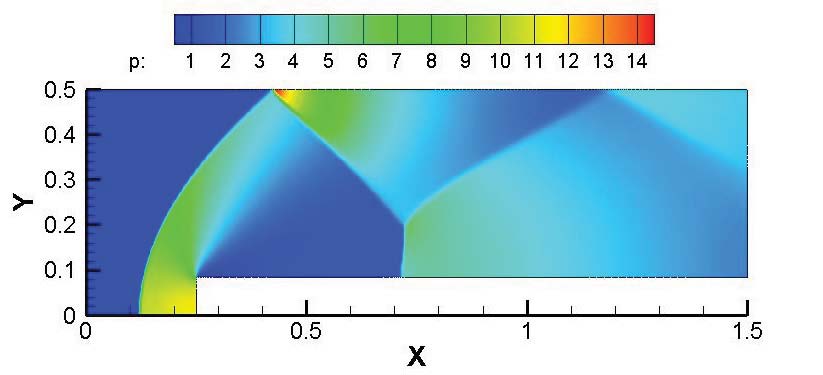}
  \caption{The pressure distribution of the flow over a forward-facing step.}
  \label{fig:Fig17}
\end{figure}

Under unstructured mesh, convergence can be achieved at rather lower computational cost with coarse mesh. But resolution should be improved by AMR. Finally, the shock wave is captured with very high accuracy. The density and pressure distributions are plotted in Fig.~\ref{fig:Fig16} and Fig.~\ref{fig:Fig17} respectively.

In CFD community, resolution for flow field depends on mesh. To resolve delicate flow structure such as shock wave, boundary layer and other region in which flow variables change greatly, cell size has to be controlled within very small value. But under structured grid, the computational cost becomes huge and many cells are distributed in unnecessary region. An efficient way to improve resolution and reduce computational cost is developing high-order scheme under unstructured hybrid mesh. Computation can be started under a coarse mesh, after convergence, adaptive mesh refinement technique is applied to improve resolution for flow field. In this way, the number of cells is much fewer than structured grid and computational cost for convergence is cut down. But in this case, shock waves and their interaction are captured accurately.

\subsection{Hypersonic flow around a circular cylinder}\label{sec:circular cylinder}

\begin{figure}
  \centering
  \includegraphics[width=0.5\textwidth]{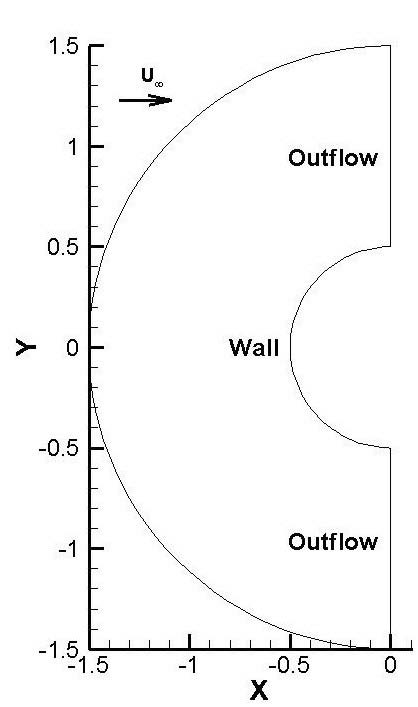}
  \caption{The computational domain for the hypersonic flow over a circular cylinder.}
  \label{fig:Fig18}
\end{figure}

A hypersonic flow passes around a unit circular cylinder. For comparison with experimental data, only half cylinder is tested. The flow condition is given as $M{a_\infty } = 8.03$, ${{\mathop{\rm Re}\nolimits} _\infty } = 183500$, and ${T_\infty } = 124.94K$. The adiabatic boundary condition is implemented on solid wall and the wall temperature is $294.44K$. The computational domain and boundary condition are given in Fig.~\ref{fig:Fig18}.

\begin{figure}
  \centering
  \includegraphics[width=0.5\textwidth]{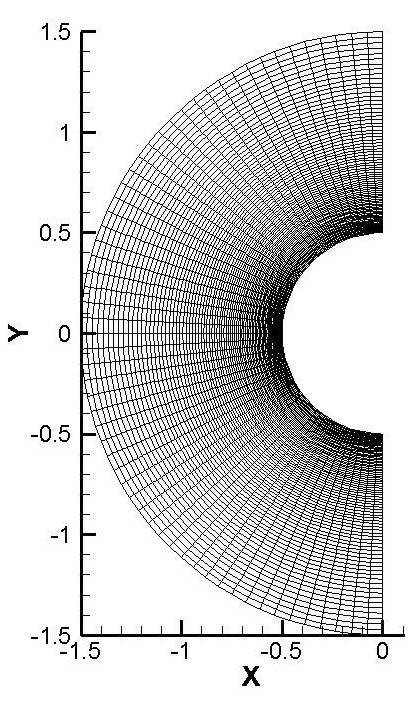}
  \caption{The initial mesh for the hypersonic flow over a circular cylinder.}
  \label{fig:Fig19}
\end{figure}

\begin{figure}
  \centering
  \includegraphics[width=0.5\textwidth]{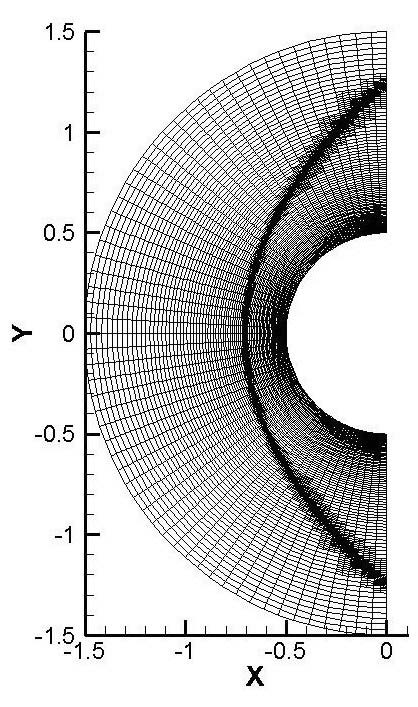}
  \caption{The refined mesh for the hypersonic flow over a circular cylinder.}
  \label{fig:Fig20}
\end{figure}

To reduce the number of mesh cells, AMR is applied to resolute the shock wave and the boundary layer. The initial mesh plotted in Fig.~\ref{fig:Fig19} contains 4800 cells. After convergence, the mesh is refined around region of shock wave and near wall. The final hybrid mesh with 14172 mesh cells is shown in Fig.~\ref{fig:Fig20}.

\begin{figure}
  \centering
  \includegraphics[width=0.5\textwidth]{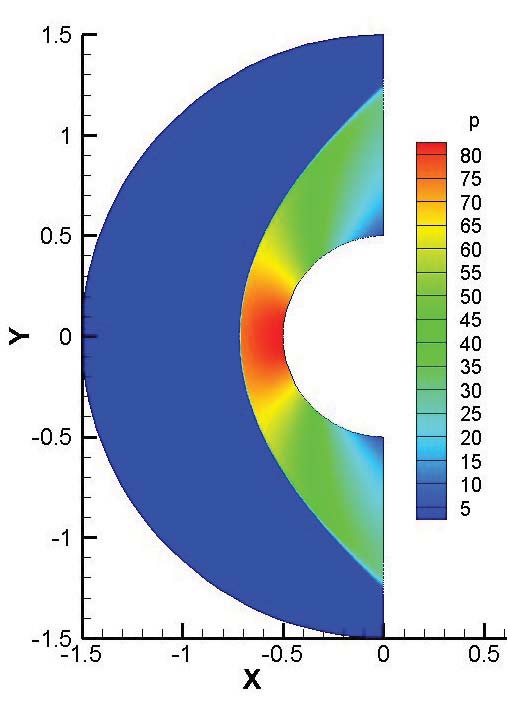}
  \caption{The pressure distribution of the hypersonic flow over a circular cylinder.}
  \label{fig:Fig21}
\end{figure}

\begin{figure}
  \centering
  \includegraphics[width=0.5\textwidth]{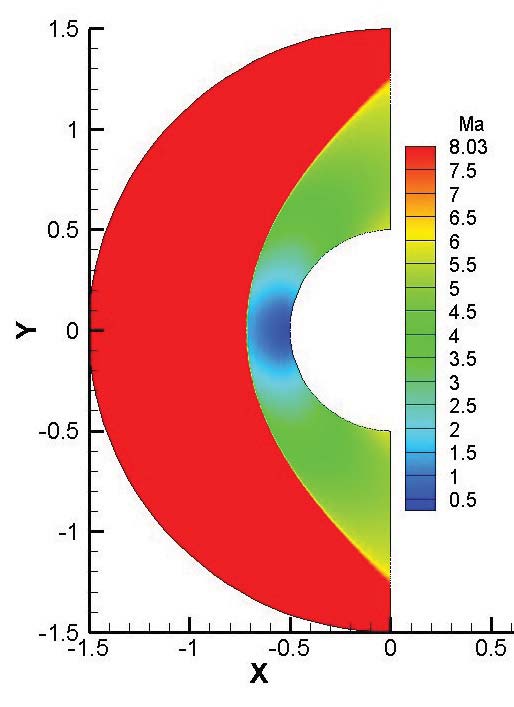}
  \caption{The Mach number distribution of the hypersonic flow over a circular cylinder.}
  \label{fig:Fig22}
\end{figure}

The pressure and Mach number distribution are given in Fig.~\ref{fig:Fig21} and Fig.~\ref{fig:Fig22}. This results are consistent with that in Ref.~\cite{Pan2016An}.

\begin{figure}
  \centering
  \includegraphics[width=0.5\textwidth]{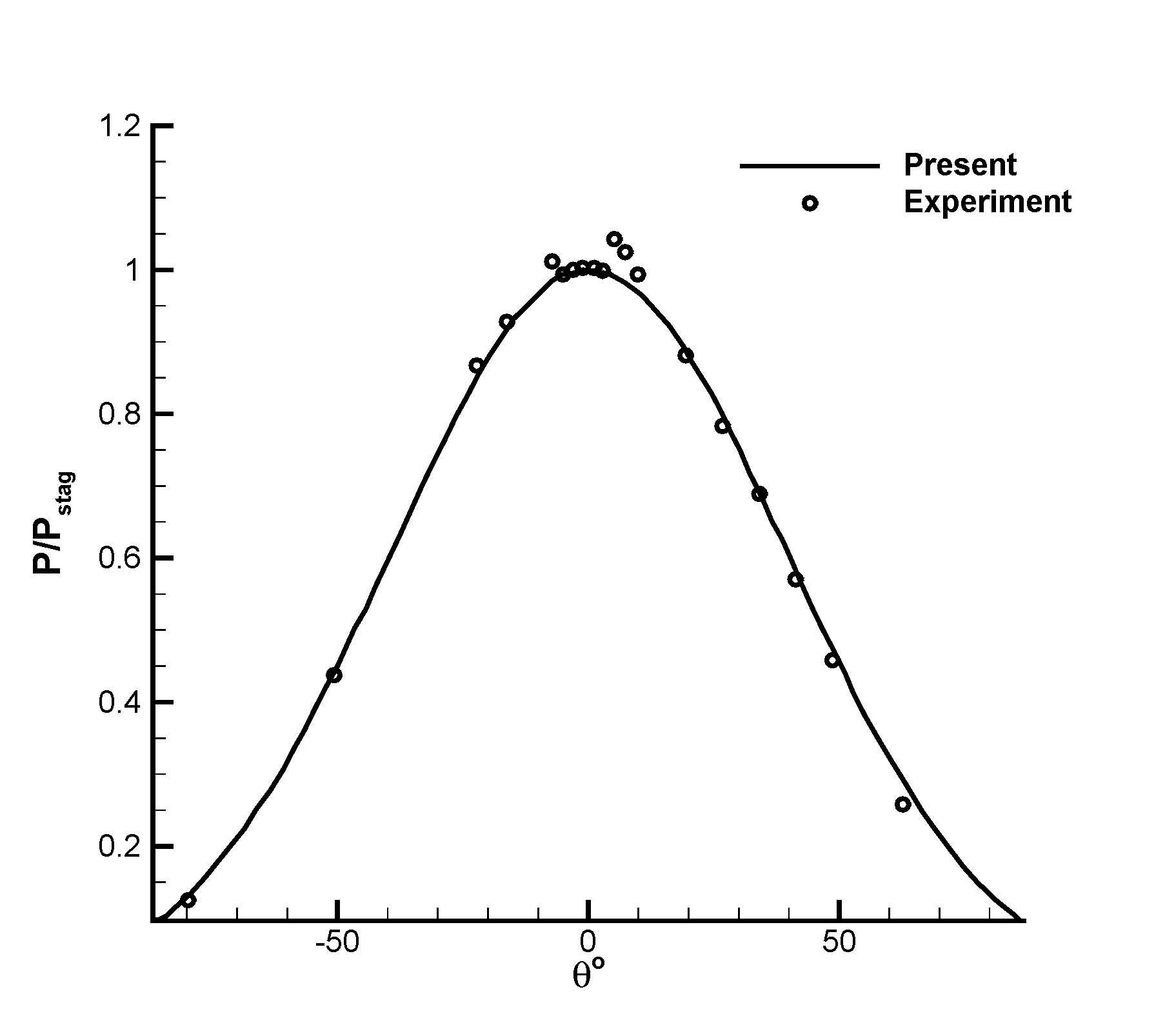}
  \caption{The pressure distribution on the surface of a circular cylinder in hypersonic flow.}
  \label{fig:Fig23}
\end{figure}

\begin{figure}
  \centering
  \includegraphics[width=0.5\textwidth]{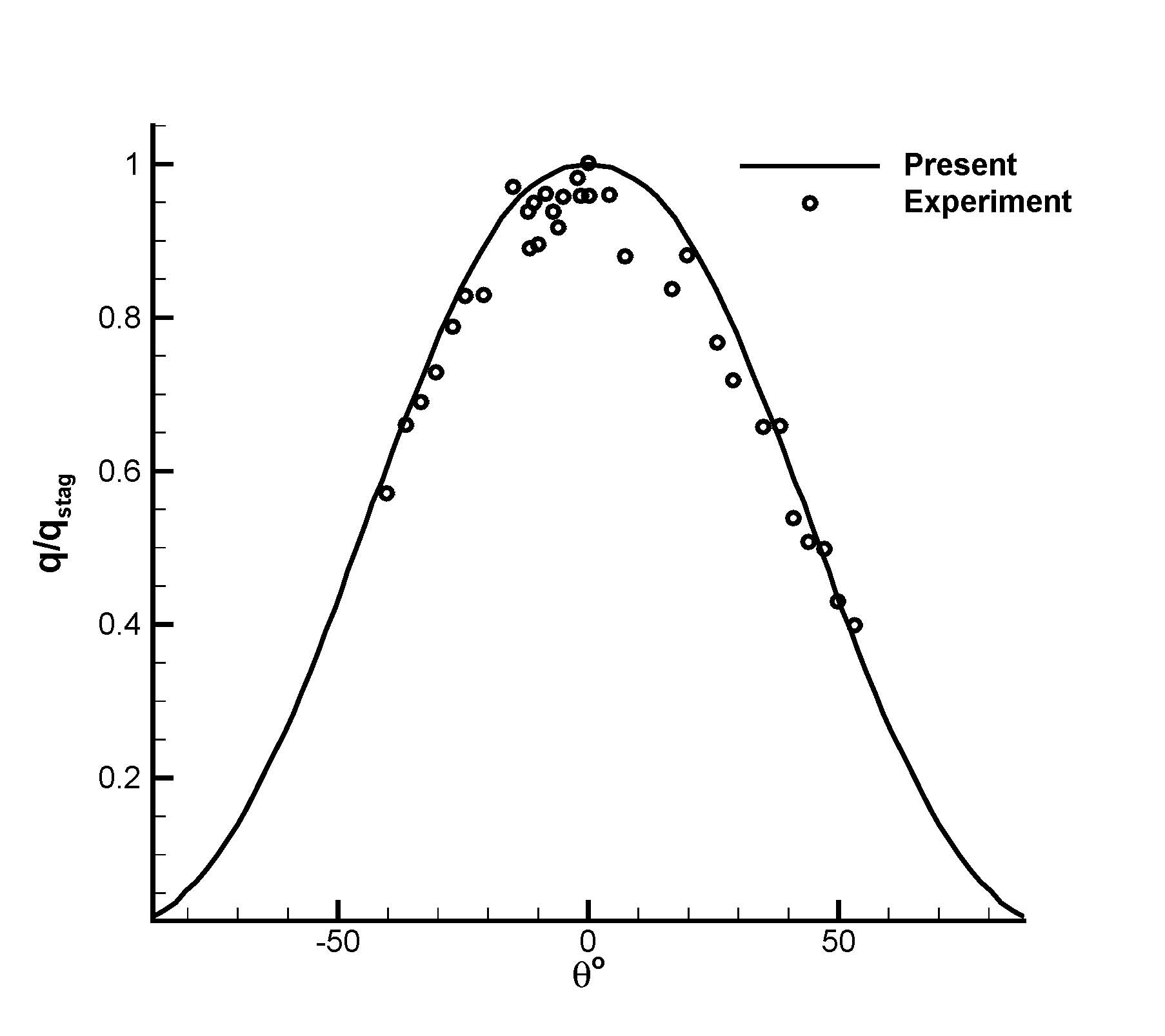}
  \caption{The heat flux distribution on the surface of a circular cylinder in hypersonic flow.}
  \label{fig:Fig24}
\end{figure}

The aerodynamic property and heat conducting is tested in this case. So the pressure and heat flux along the surface of cylinder are compared with Ref.~\cite{wieting1987experimental}. Comparison results are plotted in Fig.~\ref{fig:Fig23} and Fig.~\ref{fig:Fig24} respectively.

Computational results based on the numerical method proposed in this paper agree very well with experimental data. In this case, unstructured hybrid mesh is very flexible in implementing AMR to reduce the number of mesh cells. The carbuncle phenomenon does not appear in this hypersonic flow. The robustness of present method is also very good in reconstruction for hybrid mesh.

In this case, the hypersonic viscous flow is simulated under unstructured hybrid mesh. To resolve the shock wave, the hybrid mesh is applied in AMR around the discontinuity. The way of high-order interpolation is proved to be robust in non-manifold mesh data structure. In mesh refinement for boundary layer around surface of a cylinder, newly added vertex should be distributed according to curves of circular cylinder. So, the two-stage fourth-order GKS under unstructured mesh blended with AMR is satisfied to accurately compute incompressible and compressible flows and limits the number of cells. These advantages extend application of high-order accurate schemes to simulate flows over bodies of complex configuration.

\section{Conclusions}\label{sec:conlustion}
In this paper, a two-stage fourth-order gas-kinetic scheme based on a cubic spline interpolation under unstructured hybrid mesh is proposed for subsonic flow, supersonic flow and hypersonic flow. Instead of the four-stages Runge-Kutta conventional method for temporal evolution, the current two-stage method is more efficient in time advancing. In flux evaluation, the cubic spline interpolation is applied to reconstruct equilibrium state. To treat flow discontinuities, a hybrid WENO/spline approach is adopted to avoid spurious oscillations. This method is efficient under unstructured mesh especially when blended with AMR. To reduce the number of cells and cut down unnecessary computational cost, the advantage in dealing with non-uniform mesh which is not distributed regularly is crucial. Several cases are conducted to demonstrate the performance of the proposed scheme. Both in smooth flow and strong discontinuity, shock wave/boundary interaction, the proposed method is very robust and can give very accurate results.

\section*{Acknowledgements}
The project has been financially supported by the National Natural Science Foundation of China (Grant No. 11472219), the 111 Project of China (B17037), as well as the ATCFD Project (2015-F-016).

\clearpage
\bibliographystyle{elsarticle-num}
\bibliography{HGKS}

\begin{thebibliography}{10}
\expandafter\ifx\csname url\endcsname\relax
  \def\url#1{\texttt{#1}}\fi
\expandafter\ifx\csname urlprefix\endcsname\relax\def\urlprefix{URL }\fi
\expandafter\ifx\csname href\endcsname\relax
  \def\href#1#2{#2} \def\path#1{#1}\fi

\bibitem{Bhatnagar1954A}
P.~L. Bhatnagar, E.~P. Gross, M.~Krook, A model for collision processes in
  gases. {I. Small} amplitude processes in charged and neutral one-component
  systems, Physical Review 94~(3) (1954) 511--525.

\bibitem{Xu2001A}
K.~Xu, A gas-kinetic {BGK} scheme for the {Navier-Stokes} equations and its
  connection with artificial dissipation and {Godunov} method, Journal of
  Computational Physics 171~(1) (2001) 289--335.

\bibitem{xiong2011numerical}
S.~Xiong, C.~Zhong, C.~Zhuo, K.~Li, X.~Chen, J.~Cao, Numerical simulation of
  compressible turbulent flow via improved gas-kinetic bgk scheme,
  International Journal for Numerical Methods in Fluids 67~(12) (2011)
  1833--1847.

\bibitem{yuan2015immersed}
R.~Yuan, C.~Zhong, H.~Zhang, An immersed-boundary method based on the gas
  kinetic bgk scheme for incompressible viscous flow, Journal of Computational
  Physics 296 (2015) 184--208.

\bibitem{pan2016gas}
D.~Pan, C.~Zhong, J.~Li, C.~Zhuo, A gas-kinetic scheme for the simulation of
  turbulent flows on unstructured meshes, International Journal for Numerical
  Methods in Fluids 82~(11) (2016) 748--769.

\bibitem{li2017implementation}
J.~Li, C.~Zhong, Y.~Wang, C.~Zhuo, Implementation of dual time-stepping
  strategy of the gas-kinetic scheme for unsteady flow simulations, Physical
  Review E 95~(5) (2017) 053307.

\bibitem{Li2010A}
Q.~Li, K.~Xu, S.~Fu, A high-order gas-kinetic {Navier-Stokes} flow solver,
  Journal of Computational Physics 229~(19) (2010) 6715--6731.

\bibitem{Ren2015A}
X.~Ren, K.~Xu, S.~Wei, C.~Gu, A multi-dimensional high-order discontinuous
  {Galerkin} method based on gas kinetic theory for viscous flow computations,
  Journal of Computational Physics 292 (2015) 176--193.

\bibitem{Tang2014A}
L.~Na, H.~Tang, A high-order accurate gas-kinetic scheme for one-and
  two-dimensional flow simulation, Communications in Computational Physics
  15~(4) (2014) 911--943.

\bibitem{Lax1960Systems}
P.~Lax, B.~Wendroff, Systems of conservation laws, Communications on Pure \&
  Applied Mathematics 13~(2) (1960) 217--237.

\bibitem{Pan2016An}
L.~Pan, K.~Xu, Q.~Li, J.~Li, An efficient and accurate two-stage fourth-order
  gas-kinetic scheme for the euler and {Navier-Stokes} equations, Journal of
  Computational Physics 326 (2016) 197--221.

\bibitem{Ebeida2006Fast}
M.~Ebeida, R.~Davis, Fast adaptive hybrid mesh generation based on quad-tree
  decomposition, in: Fluid Dynamics Conference and Exhibit, 2006.

\bibitem{Pan2016A}
D.~Pan, C.~Zhong, J.~Li, C.~Zhuo, A gas-kinetic scheme for the simulation of
  turbulent flows on unstructured meshes, International Journal for Numerical
  Methods in Fluids 82~(11) (2016) 748--769.

\bibitem{Liu2016A}
Y.~Liu, W.~Zhang, Y.~Jiang, Z.~Ye, A high-order finite volume method on
  unstructured grids using {RBF} reconstruction, Computers \& Mathematics with
  Applications 72~(4) (2016) 1096--1117.

\bibitem{Liu2017Accuracy}
Y.~Liu, W.~Zhang, Accuracy preserving limiter for the high-order finite volume
  method on unstructured grids, Computers \& Fluids 149 (2017) 88--99.

\bibitem{Gamet1999Compact}
L.~Gamet, F.~Ducros, F.~Nicoud, T.~Poinsot, Compact finite difference schemes
  on non-uniform meshes. {Application} to direct numerical simulations of
  compressible flows, International Journal for Numerical Methods in Fluids
  29~(2) (1999) 159--191.

\bibitem{Burbeau2002Simulation}
A.~Burbeau, P.~Sagaut, Simulation of a viscous compressible flow past a
  circular cylinder with high-order discontinuous {Galerkin} methods, Computers
  \& Fluids 31~(8) (2002) 867--889.

\bibitem{Wang2015An}
Q.~Wang, Y.~X. Ren, An accurate and robust finite volume scheme based on the
  spline interpolation for solving the {Euler and Navier-Stokes} equations on
  non-uniform curvilinear grids, Journal of Computational Physics 284 (2015)
  648--667.

\bibitem{schoenberg1988contributions}
I.~J. Schoenberg, Contributions to the problem of approximation of equidistant
  data by analytic functions, in: IJ Schoenberg Selected Papers, Springer,
  1988, pp. 3--57.

\bibitem{Rubin1975Higher}
S.~G. Rubin, P.~K. Khosla, Higher-order numerical solutions using cubic
  splines, AIAA Journal 14~(7) (1975) 851--858.

\bibitem{Li2015A}
J.~Li, Z.~Du, A two-stage fourth order time-accurate discretization for
  {Lax-Wendroff} type flow solvers. {I. Hyperbolic} conservation laws, SIAM
  Journal on Scientific Computing 38~(5) (2016) 3046--3069.

\bibitem{Ren2003A}
Y.~X. Ren, M.~Liu, H.~Zhang, A characteristic-wise hybrid {compact-WENO} scheme
  for solving hyperbolic conservation laws, Journal of Computational Physics
  192~(2) (2003) 365--386.

\bibitem{Shu1989Efficient}
C.~W. Shu, S.~Osher, Efficient implementation of essentially non-oscillatory
  shock-capturing schemes, Journal of Computational Physics 77~(2) (1989)
  439--471.

\bibitem{Jiang1996Efficient}
G.~S. Jiang, C.~W. Shu, Efficient implementation of weighted {ENO} schemes,
  Academic Press Professional, Inc., 1996.

\bibitem{Alexander2002Solution}
K.~Alexander, E.~Tadmor, Solution of two-dimensional {Riemann} problems for gas
  dynamics without {Riemann} problem solvers, Numerical Methods for Partial
  Differential Equations 18~(5) (2002) 584--608.

\bibitem{Daru2009Numerical}
V.~Daru, C.~Tenaud, Numerical simulation of the viscous shock tube problem by
  using a high resolution monotonicity-preserving scheme, Computers \& Fluids
  38~(3) (2009) 664--676.

\bibitem{Kim2005Accurate}
K.~H. Kim, C.~Kim, Accurate, efficient and monotonic numerical methods for
  multi-dimensional compressible flows, Journal of Computational Physics
  208~(2) (2005) 527--569.

\bibitem{Daru2000Evaluation}
V.~Daru, C.~Tenaud, Evaluation of {TVD} high resolution schemes for unsteady
  viscous shocked flows, Computers \& Fluids 30~(1) (2000) 89--113.

\bibitem{wieting1987experimental}
A.~R. Wieting, M.~S. Holden, Experimental study of shock wave interference
  heating on a cylindrical leading edge at mach 6 and 8, in: AIAA, 22nd
  Thermophysics Conference, 1987.

\end{thebibliography}

\clearpage

\clearpage

\end{document}